\documentclass[aps, superscriptaddress, prd, amsmath, floats, eqsecnum, floatfix, onecolumn, nofootinbib,10pt]{revtex4-2}

\usepackage{graphicx}
\usepackage[format=plain, font=small, labelfont=bf, justification=RaggedRight]{caption}
\usepackage[format=plain, font=small, labelfont=bf, justification=RaggedRight]{subcaption}
\usepackage{color}
\usepackage{latexsym}
\usepackage{slashed}
\usepackage{amssymb}
\usepackage{orcidlink}				 

\renewcommand{\thesection}{\arabic{section}}
\renewcommand{\thesubsection}{\thesection.\arabic{subsection}}

\makeatletter
\def\p@subsection{}
\def\p@subsubsection{}
\makeatother

\usepackage[titletoc]{appendix}

\newcommand{\rd}{\mathrm{d}}	
\usepackage{mathrsfs}			
\usepackage{txfonts}			

\newcommand{\be}{\begin{eqnarray}}
\newcommand{\ee}{\end{eqnarray}}

\newcommand{\bea}{\begin{eqnarray}}
\newcommand{\eea}{\end{eqnarray}}

\newcommand{\beq}{\begin{equation}}
\newcommand{\eeq}{\end{equation}}

\newcommand{\eq}[1]{(\ref{#1})}
\newcommand{\n}[1]{\label{#1}}

\newcommand{\al}{\alpha}
\newcommand\ga{\gamma}

\newcommand\de{\delta}

\renewcommand\th{\theta}

\newcommand\la{\lambda}

\newcommand\si{\sigma}

\newcommand\ph{\varphi}

\newcommand\lap{\Delta}
\newcommand\na{\nabla}

\makeatletter
\def\l@subsubsection#1#2{} 	
\makeatother

\begin{document}

\title{
On effective models of regular black holes inspired by higher-derivative and nonlocal gravity
}

\author{Tib\'{e}rio de Paula Netto\orcidlink{0000-0001-9619-2358}}
\email{tiberio.netto@ufjf.br}
\affiliation{
{\small Departamento de F\'{\i}sica,  ICE, Universidade Federal de Juiz de Fora,
Juiz de Fora,  36036-900,  MG,  Brazil}
}

\author{Breno L. Giacchini\orcidlink{0000-0002-6449-7189}}
\email{breno.giacchini@matfyz.cuni.cz}
\affiliation{
{\small Institute of Theoretical Physics, Faculty of Mathematics and Physics, Charles University, V Hole{\v s}ovi{\v c}k{\'a}ch 2, 180 00 Prague 8, Czech Republic}
}
\affiliation{
{\small Departamento de F\'{\i}sica,  ICE, Universidade Federal de Juiz de Fora,
Juiz de Fora,  36036-900,  MG,  Brazil}
}

\author{Nicol\`o Burzill\`a\orcidlink{0000-0002-7740-0200}}
\email{burzilla@roma2.infn.it}
\affiliation{
{\small INFN Sezione di Roma Tor Vergata, 00133 Roma, Italy}
}
\affiliation{
{\small Department of Physics, Southern University of Science and Technology, Shenzhen 518055, China}
}

\author{Leonardo Modesto}
\email{leonardo.modesto@unica.it}
\affiliation{
{\small Dipartimento di Fisica, Universit\`a di Cagliari, Cittadella Universitaria, 09042 Monserrato, Italy}
}
\affiliation{
{\small Department of Physics, Southern University of Science and Technology, Shenzhen 518055, China}
}


\begin{abstract} 
\vspace{0.1cm}
\noindent
In this work we study static spherically symmetric solutions of effective field equations related to local and nonlocal higher-derivative gravity models, based on 
their associated effective delta sources. This procedure has been applied to generate modifications of the Schwarzschild geometry in several contexts (e.g., modified gravity, string theory, noncommutative geometry, generalized uncertainty principle scenarios), but a general analysis of the possible equations of state and their influence on the solutions was still lacking. Here, we aim to fill this gap in the literature and investigate whether these metrics might be able to reproduce features of the solutions of higher-derivative gravity models.  In particular, we present an equation of state such that the solution matches the Newtonian-limit one in both regimes of large and small~$r$. A significant part of the work is dedicated to studying the curvature regularity of the solutions and the comparison with the linearized solutions. Explicit metrics are presented for effective sources originating from local and nonlocal models. The results obtained here might be regarded as possible links between the previous research on linearized higher-derivative gravity and the solutions of the nonlinear complete field equations, which remain unknown at the moment.
\end{abstract}

\maketitle
\noindent

\vspace{-0.9cm}

\tableofcontents

\section{Introduction}
\label{introduction}
\label{Sec1}

Higher-derivative gravitational theories attract considerable attention due to their numerous applications in quantum gravity. Indeed, the simple integration of vacuum bubble diagrams of matter fields in curved spacetime produces curvature-squared terms in the effective action~\cite{UtDW}, while the inclusion of the same type of structures in the classical gravitational action of general relativity renders the quantum theory renormalizable~\cite{Stelle77}. Adding more derivatives to the action enhances the ultraviolet (UV) behavior of the quantum theory even more. In fact, terms quadratic in curvatures and polynomial in derivatives can make the theory superrenormalizable~\cite{AsoreyLopezShapiro}. Theories with an infinite number of derivatives may also be (super)renormalizable depending on the choice of the action nonlocal form factor~\cite{Krasnikov,Kuzmin,Tomboulis,Modesto12}.

The literature on models with more than four metric derivatives in the gravitational action has increased substantially in the last decade. This is motivated by the possibility of solving the contradiction between renormalizability and unitarity present in the original formulation of fourth-derivative gravity. Even though the quantum properties of these models are well understood nowadays, the same cannot be said about their classical solutions.

The aim of the present work is to systematically study a procedure for generating modifications of the Schwarzschild geometry inspired by higher-derivative gravity models and discuss the main features of the resulting metrics. Although these metrics are not solutions to the complete field equations of those models, they might reproduce characteristic properties of the fundamental solutions.

The underlying higher-derivative models we focus on belong to a large class of theories quadratic in curvatures, described by the action
\beq
\n{action-0}
S = \frac{1}{16 \pi G} \int \rd^4 x \sqrt{-g} \left[   R  +  R_{\mu\nu} F_1 (\Box) R^{\mu\nu} +  R  F_2 (\Box)   R  \right] 
,
\eeq 
where $G$ is the Newton constant and $F_{1,2}(\Box)$ are form factors, analytic functions of the d'Alembert operator subjected to the constraint\footnote{This assumption ensures that the theory contains higher derivatives both in the spin-2 and spin-0 sectors, which is a necessary condition for renormalizability. Throughout this paper, unless otherwise stated, we always assume that $F_1 \neq - 3 F_2$. Moreover, it is always assumed that $F_{1,2}$ are such that the theories do not contain tachyons.} $F_1 \neq - 3 F_2$. In what follows we list some popular choices of form factors and their respective quantum properties regarding renormalizability\footnote{In general, for renormalization considerations, the action~\eq{action-0} should also include the cosmological constant and two terms that do not affect the classical equations of motion, namely, $\Box R$ and the Gauss--Bonnet topological invariant.}  and the presence of ghost-like degrees of freedom:
\begin{itemize}
\item[(i)] $F_1 (z) = F_2 (z) \equiv 0$ corresponds to the trivial case where we recover general relativity, which is a unitary but nonrenormalizable theory~\cite{hove,dene,GorSag1,GorSag2}.

\item[(ii)] If $F_{1,2} (z)$ are nonzero constants we meet the fourth-derivative theory, also known as Stelle gravity~\cite{Stelle77}. This model is renormalizable, but it is nonunitary due to the presence of a spin-2 massive ghost (if quantized using the standard techniques of quantum field theory). Recent proposals for solving the problem of the ghost in the context of fourth-derivative gravity can be found, e.g., in~\cite{Bender:2007wu,Bender:2008gh,Salvio:2015gsi,Anselmi:2017ygm,Donoghue:2019fcb}.

\item[(iii)] If $F_{1,2} (z)$ are polynomials of degree $N \geqslant 1$ the theory is superrenormalizable~\cite{AsoreyLopezShapiro}. 
The unitarity problem in these models can be tackled by choosing form factors such that all ghost-like degrees of freedom have complex masses and performing the quantization {\it \`{a} la} Lee--Wick~\cite{LW1,LW2,CLOP,ModestoShapiro16,Modesto16,AnselmiPiva2}. This subclass of polynomial-derivative gravity is also known as Lee--Wick gravity~\cite{ModestoShapiro16,Modesto16}.

\item[(iv)] For nonlocal form factors of the type
\beq
\n{nonlocal-ff}
F_{2} (\Box) = \frac{e^{H_{2}(\Box)} - 1}{\Box} , \qquad  F_1(\Box) = - \frac{e^{H_{1}(\Box)} - 1}{\Box} - 3 F_2(\Box) 
,
\eeq
where $H_{1,2}(z)$ are entire functions, the propagator of the theory only contains the massless pole corresponding to the graviton. Thus, these models are ghost-free at tree level and, depending on the choice of the functions $H_{1,2}(z)$, they can be (super)renormalizable~\cite{Krasnikov,Kuzmin,Tomboulis,Modesto12}.
\end{itemize}

Quadratic structures of the type $R_{\mu\nu\al\beta} F_3 (\Box) R^{\mu\nu\al\beta}$ and 
$O(R^3_{\cdots})$ terms 
could also be included in the action~\eq{action-0} without substantially changing the general results concerning (super)renormalizability. Since the latter type of terms only affect the interaction vertices (and not the propagator), provided that they do not contain more derivatives than the curvature-quadratic part, they cannot spoil the renormalizability. On the contrary, by a judicious choice of these structures, the theories described in (iii) and (iv) may even become finite at quantum level~\cite{AsoreyLopezShapiro,Modesto:2014lga}. As for the term quadratic in the Riemann tensor, it modifies both propagator and vertices in a {very similar way to} the structures in~\eq{action-0}, and the analysis of renormalizability can be performed in an analogous manner (see, e.g.,~\cite{AsoreyLopezShapiro}). Nonetheless, the space of classical solutions can be significantly affected by the presence of these two kinds of terms in the action. For the sake of simplicity, we do not consider such terms in the present work; instead, we focus only on a particular class of the model~\eq{action-0}.

The systematic search and study of spherically symmetric solutions in theories quadratic in curvature  dates back to the 1970s~\cite{Wynne73,Stelle78,FroVilk,FroVilkMG} (see also~\cite{PS66,Havas77} and references therein for earlier considerations). Some authors have argued that the Schwarzschild solution is the only static, spherically symmetric, and asymptotically flat solution of fourth-derivative gravity in four dimensions (see, e.g.,~\cite{Frolov:2009qu,Nelson:2010ig}). However, in~\cite{Lu:2015cqa} it was demonstrated the existence of other solutions that deviate from the Schwarzschild one. After that, the properties of static spherically symmetric solutions in Stelle gravity became an active research topic~\cite{Stelle15PRD,Cai:2015fia,Feng:2015sbw,Lin:2016kip,Holdom:2016nek,Holdom:2022zzo,Holdom:2015kbf,Lu:2017kzi,Kokkotas:2017zwt,Goldstein:2017rxn,Podolsky:2018pfe,Svarc:2018coe,Holdom:2019ouz,Salvio:2019llz,Aydemir:2020pao,Bonanno:2019rsq,Konoplya:2019ppy,Bonanno:2021zoy,Daas:2022iid,Silveravalle:2022wij}. 
The main results can be summarized very briefly as follows: In the complete fourth-derivative gravity, the solution that couples to a delta source 
contains a naked singularity; other solutions, corresponding to regular metrics, could describe wormholes. Event horizons may be present only in the particular case of Einstein--Weyl gravity.

With respect to the models with more than four derivatives in the action or with nonlocal form factors, most of the solutions found in the literature are in the linear approximation~\cite{Newton-MNS,Newton-BLG,BreTib1,BreTib2,Nos4der,Nos6der,Frolov:Poly,Frolov:Exp,Boos:2018bxf,Boos:2020ccj,Buoninfante:2020qud,Buoninfante:2018xif,Kolar:2020bpo,Heredia:2021pxp,Accioly:2016qeb,Head-On,Tseytlin:1995uq,SiegelEtAl1,SiegelEtAl2,SiegelEtAl3,Edholm:2016hbt,Kilicarslan:2018yxd,Buoninfante:2018xiw,Buoninfante:2018rlq}.\footnote{
See also, e.g.,~\cite{Kilicarslan:2019njc,Dengiz:2020xbu,Kolar:2021rfl,Kolar:2021uiu,Kolar:2023gqi} and references therein for considerations of exact solutions, and~\cite{Frolov:2023jvi} for a recent discussion on nonlocal modifications of the Kerr metric.}
The study of the theory in the Newtonian limit is certainly interesting, for it gives the large-distance behavior of the metric---which should coincide with the asymptotic limit of the  non-approximated solution. The small-distance behavior of the linearized solutions has also been the subject of intensive studies, and it was shown that the modified Newtonian potential is bounded and the curvature scalars are regular in most of the higher-derivative theories. To be more precise,  all models of the type~\eq{action-0} with nonzero $F_1(z)$ and $F_2(z)$  have a finite Newtonian potential~\cite{Newton-MNS,Newton-BLG}; while if $F_{1,2} (z)$ are nontrivial polynomials (i.e., the action has six or more derivatives) the model admits a regular Newtonian-limit solution, without singularities in the curvature invariants~\cite{BreTib1}. The generalization of these statements to nonlocal theories can be found in~\cite{Nos6der}. Basically, what is important for the absence of singularities is the UV behavior of the form factors~\cite{BreTib2,Nos6der}. If $H_{1,2}(z)$ in~\eq{nonlocal-ff} is such that $F_{1,2}(z)$ behave like a polynomial of degree $N$ at high energies (like in the cases of Kuz'min and Tomboulis form factors~\cite{Kuzmin,Tomboulis}), the nonlocal theory reproduces the regularity properties of the corresponding (local) polynomial model. If $F_{1,2}(z)$ grow faster than any polynomial (for large $z$), not only the curvature invariants are regular at $r=0$, but so are all the invariants that are polynomial in derivatives of curvature tensors (collectively known as {\it curvature-derivative invariants}~\cite{Nos6der,Giacchini:2021pmr,BreTibLiv}). For the detailed consideration of curvature-derivative invariants, see~\cite{Nos6der,Giacchini:2021pmr}.

Even though the linear version of the theory does not seem to be the appropriate setting to investigate whether higher derivatives could resolve the singularity, it has been speculated that the solution of the full nonlinear theory might be regular as well.  It was even ventilated that close to $r=0$ the gravitational field may be weak enough so that one could trust the linearized solution in a small region around the origin~\cite{Li:2015bqa}---or, under certain circumstances, from far away up to the origin~\cite{Buoninfante:2018rlq}.

To the best of our knowledge, static spherically symmetric solutions of the full nonlinear equations of motion of curvature-quadratic gravity theories with more than four derivatives of the metric have been considered only in~\cite{Holdom:2002xy}, based on the Frobenius method and numerical analysis of the field equations in models with  up to ten derivatives.\footnote{After the completion of the present work, aspects of solutions of such sixth-derivative gravity models were studied in~\cite{Giacchini:2024exc,Pawlowski:2023dda,Daas:2024pxs}.} The conclusion of this study can be summarized as follows:
\begin{itemize}
\item[(I)] All the solutions found  are regular at $r=0$.
\item[(II)] There is a critical mass $M_{\rm c}$ for the formation of a black hole. If $M < M_{\rm c}$ 
the solution does not have an event horizon. 
On the other hand, for $M > M_{\rm c} $ the solution possesses more than one horizon. 
\end{itemize}
Moreover, it was speculated whether some appealing properties of the regular solutions of fourth-derivative gravity could be shared by their higher-derivative counterparts, namely:
\begin{itemize}
\item[(III)] Nonlinearities are suppressed sufficiently close to $r=0$; the solutions behave like the weak-field ones. 
\item[(IV)] The gravitational field is strong near the position of the horizons; in these regions, there is a transition between the linear and nonlinear regimes.
\end{itemize}

Last but not least, owing to the complexity of the equations of motion of higher-derivative gravity, the works~\cite{Modesto12,Modesto:2010uh,Bambi:2016wmo,Zhang14} considered a truncated form of the field equations associated to specific models defined by the action 
\beq
\label{action}
S = \frac{1}{16 \pi G} \int \rd^4 x \sqrt{-g} \left[ R + G_{\mu\nu} F(\Box) R^{\mu\nu} \right]
,
\eeq
where $G_{\mu\nu}$ is the Einstein tensor.
The action \eq{action} corresponds to a particular class of theories of the type~\eq{action-0}, namely, the ones whose form factors satisfy the relation 
\beq
F_1 (\Box) = - 2 F_2 (\Box) \equiv F(\Box). 
\eeq
The truncation of the equations of motion is sometimes referred to as the {\it propagator approximation}, or {\it smeared delta source approximation}.
Since the present work is devoted precisely to solutions of this type, the remaining part of this introductory section describes the procedure and the underlying assumptions.

Simply put, the procedure {consists in} a curved-spacetime generalization of the method for calculating the Newtonian potential of a pointlike source in higher-derivative models. It can be shown (see, e.g.,~\cite{BreTibLiv} or Sec.~\ref{Sec5} below) that in the Newtonian limit, the solution of the equations of motion associated to~\eq{action} {\it only depends on one potential}, $\ph (r)$. In fact, the linear limit metric written in Schwarzschild coordinates is given by
\beq
\n{metric-New}
\rd s^2 = - \left[1 + 2 \ph(r) \right] \rd t^2 + \left[ 1 + 2 r \ph'(r) \right] \rd r^2 + r^2 \rd \Omega^2
,
\eeq
where $\rd \Omega^2 = \rd \th^2 + \sin^2 \th \rd \phi^2$ 
and primes denote differentiation with respect to the coordinate $r$. 
The potential for a pointlike source is thus the solution of the modified Poisson equation (see, e.g.,~\cite{BreTib1,BreTibLiv})
\beq
\n{new-pot-eq-1}
f (\lap) \lap \ph = 4 \pi G \rho
,
\quad \quad
\rho (\vec{r}) = M \delta^{(3)} (\vec{r}),
\eeq
where
\beq
\n{fzinho}
f (z) = 1 + z F (z).
\eeq
In an equivalent form, $\ph(r)$ can be obtained by solving the equation
\beq
\n{new-pot-eq-2}
\lap \ph = 4 \pi G \rho_{\rm eff}
,
\eeq 
where
\beq
\n{effsource}
\rho_{\rm eff}(r) = \frac{M}{2 \pi ^2 r} 
\int_0^\infty \rd k \, \frac{ k \sin (kr)}{ f(-k^2) }
\eeq
is the {\it smeared delta source}, which follows from the inversion of the operator $f(\lap)$ in \eq{new-pot-eq-1}.\footnote{Some conditions must be imposed on the function $f(z)$ for the integral~\eq{effsource} to be well defined. Namely, we assume that $f(-k^2)>0$ for $k\in\mathbb{R}$, $f(0)=1$, and that, if $f(-k^2)$ is not trivial, it diverges at least as fast as $k^2$ for $k \to \infty$. These conditions represent constraints on the form factor $F(\Box)$. In particular, the former one avoids tachyonic poles in the propagator, and the latter ones restrict the type of weakly nonlocality (they are immediately satisfied in local models); see~\cite{BreTib2,BreTibLiv} for details.} Therefore, Eq.~\eq{new-pot-eq-1} can be cast as a standard Poisson equation~\eq{new-pot-eq-2} with the effective delta source~\eq{effsource} (see, e.g.,~\cite{BreTib2,BreTibLiv}). Hence, in flat spacetime, the effect of the higher derivatives may be regarded as inducing a smearing of the original delta source, through the nonconstant function $f(z)$ in the integrand of~\eq{effsource}.

The main idea, now, is to obtain a curved-space generalization of~\eq{new-pot-eq-2} to static spherically symmetric spacetimes in the form
\beq
\label{nlEE2}
G^{\mu} {}_{\nu} = 8 \pi G \, \tilde{T}^{\mu} {}_{\nu},
\eeq
where $\tilde{T}^{\mu} {}_{\nu}$ is an effective energy-momentum tensor reproducing the effects of the higher derivatives 
with 
\beq
\n{Teff-00}
\tilde{T}^{t} {}_{t} = - \rho_{\rm eff}
,
\eeq
being $\rho_{\rm eff}$ the smeared delta source~\eq{effsource}.
In this way, like Eq.~\eqref{new-pot-eq-2} yields the modified Newtonian potential, the metrics which follow from~\eq{nlEE2} can be interpreted as higher-derivative modifications of Schwarzschild, through the function $f(z)$. In fact, if the higher derivatives are switched off, $f(z) \equiv 1$ and Eq.~\eq{nlEE2} results in the Schwarzschild solution.

The formula~\eq{nlEE2} can also be seen as a truncation of the full equations of motion for the theory~\eq{action} sourced by a pointlike source. Indeed, owing to the specific form of the higher-derivative structure in~\eq{action}, it is possible to factor the operator $f(\Box)$ together with the Einstein tensor in the equations of motion, namely,
\beq
\n{cu}
f(\Box) G^{\mu} {}_{\nu} + O(R^2_{\cdots}) = 8 \pi G \, T^{\mu} {}_{\nu}
,
\eeq
where $T^{\mu} {}_{\nu}$ is the matter usual energy-momentum tensor. Therefore, by discarding the terms $O(R^2_{\cdots})$ of higher order in curvatures in~\eq{cu} and taking the d'Alembertian in $f(\Box)$ 
in the flat-spacetime form, from the inversion of the operator $f(\Box)$ one can recast~\eq{cu} into~\eq{nlEE2}~\cite{Modesto12, Modesto:2010uh, Li:2015bqa,Bambi:2013gva, Bambi:2016wmo, Bambi:2016uda, Zhang14, Modesto:2012ys}. 
Although frequently used in the literature, this procedure of truncating the d'Alembertian (to obtain the invertible form that does not depend on the metric) is seldom discussed.
In this respect, it was argued that these truncations of the original equation~\eq{cu} might be compensated by imposing the conservation of the effective energy-momentum tensor, i.e., $\na_\mu \tilde{T}^{\mu} {}_{\nu} = 0$ (see, e.g.,~\cite{Modesto12}). This leads to the introduction of effective radial ($p_r$) and tangential ($p_\th$) pressures in the effective energy-momentum tensor,
\beq
\n{effT}
\tilde{T}^{\mu} {}_\nu = \text{diag}(-\rho_{\rm eff} , p_r , p_\th , p_\th ) .
\eeq
As we show in the present work, although the spacetime configurations constructed in this framework are not likely the solutions of the complete field equations, they reproduce relevant aspects of the findings of~\cite{Holdom:2002xy}.

In the present work, we study static spherically symmetric solutions to the system of equations~\eq{nlEE2}, with the general line element
\beq
\label{metric}
\rd s^2 = - A(r) e^{B(r)} \rd t^2 + \frac{\rd r^2}{A(r)} + r^2 \rd \Omega^2
,
\eeq
where $A(r)$ and $B(r)$ are two functions to be determined. Differently from previous works~\cite{Modesto:2010uh,Bambi:2016wmo}, here we are interested in solutions with a nontrivial shift function, $B(r) \neq 0$. 
The motivation comes from the observation that the identity $\ph(r) = - r \ph'(r)$ (which is valid in Einstein gravity) for the Newtonian potential does not hold in general for higher-derivative models, thus to reproduce the weak-field behavior \eq{metric-New} we need a metric with $g_{tt} \neq - g^{rr}$.\footnote{See~\cite{Daas:2024pxs,Giacchini:2024exc,Bueno:2017sui} for  further evidences that solutions in higher-derivative gravity models might require a nontrivial shift function.}

We mainly focus on form factors that grow at least as fast as $f(z) \sim z^2$ for large $z$. This corresponds to the smeared delta sources associated to models whose actions have at least six derivatives of the metric tensor [see~\eq{action} and~\eq{fzinho}]. In the main part of this paper, we show that under some reasonable assumptions, the modified Schwarzschild metric that follows from the simple equation~\eq{nlEE2} can reproduce the properties (I)--(IV) 
of the static spherically symmetric solutions of~\cite{Holdom:2002xy}. By imposing some physical requirements on the spatial part of $\tilde{T}^{\mu} {}_{\nu}$, we also show that for all form factors such that $f(z) \sim z^2$ or faster for large $z$, the solutions of~\eq{nlEE2} have curvature invariants without singularities.
This result is at the core of a procedure to generate a large amount of regular black hole metrics.\footnote{Several definitions of ``spacetime regularity'' exist in the literature, e.g., referring to the regularity of metric components or Christoffel symbols in a given coordinate chart, to the regularity of a set of curvature invariants, or to the geodesic completeness of the spacetime (see, e.g.,~\cite{Frolov:2016pav,Bejarano:2017fgz,Carballo-Rubio:2019fnb,Berry:2021hos} and references therein). 
As a matter of convention, in this work, we characterize a regular black hole by the absence of singularities in its Kretschmann scalar $R_{\mu\nu\al\beta} R^{\mu\nu\al\beta}$. For static and spherically symmetric geometries, this also implies in the regularity of all the curvature invariants constructed contracting an arbitrary number of Riemann and metric tensors~\cite{Bronnikov:2012wsj}. An extended definition of regularity, which also considers the absence of singularities in curvature invariants containing covariant derivatives of curvatures, will be discussed in Sec.~\ref{Sec6}.
}

It is worth pointing out that the metrics we obtain here, regarded as higher-derivative modifications of Schwarzschild, are similar in spirit to the nonlocal modifications of the Kerr and Schwarzschild metrics recently studied in~\cite{Frolov:2023jvi}. In fact, they are not solutions of higher-derivative gravity models, but encode higher-derivative modifications at equations whose original forms would result in the Schwarzschild metric. Our procedure, however, differs from the one of~\cite{Frolov:2023jvi}; for instance, their modification of the Schwarzschild metric takes the form $A(r) = 1 + 2\varphi(r)$ and $B\equiv 0$ (in our notation).

Finally, let us mention that the formulation of the problem in terms of the effective field equations~\eq{nlEE2} 
with smeared delta sources 
is standard in several other approaches to quantum gravity such as noncommutative geometry, generalized
uncertainty principle models, and string theory (see, e.g.,~\cite{Cadoni:2022chn,Cadoni:2023nrm,Akil:2022coa,Nicolini:2005vd,dirty,Mureika:2010je,Gaete:2010sp,Nicolini:2012eu,Isi:2013cxa,Tseytlin:1995uq,Nicolini:2019irw,Knipfer:2019pgi}). Therefore, the detailed study of the solutions of Einstein's equation~\eq{nlEE2} with the effective source~\eq{effsource}
is relevant not only for higher-derivative gravities but also for alternative models of quantum gravity.
      
The paper is organized as follows. In Sec.~\ref{Sec2} we present the equations for the metric components $A(r)$ and $B(r)$, which follows from the effective field equations~\eq{nlEE2}, and obtain their formal solutions depending on a generic form factor $f(z)$. Then, a brief review of the linearized field equations and their solutions is carried out in Sec.~\ref{Sec5}. In Sec.~\ref{Sec6} we discuss the curvature regularity of the solutions of the effective field equations on general grounds, i.e., without particularizing to specific form factors. We also discuss the effect of the nonlinearity of the field equations on the regularity of the solutions. From Sec.~\ref{Sec3}, the considerations start to become model-dependent. In that section, we study the general properties of the solution for $A(r)$, showing how 
the form factor $f(z)$ can define the regularity of $A(r)$ and the possible structure of horizons of the metric.
The similar analysis for the function $B(r)$ is carried out in Sec.~\ref{Sec4}, where we discuss several choices of equations of state for the effective pressures and their influence on the solutions. In particular, in Sec.~\ref{Sec4.4} we introduce an equation of state such that the solution of the effective field equations interpolates between a regular core, for small $r$, and the Newtonian-limit solution at large $r$. 
Examples where the scenarios described in the previous sections can be explicitly visualized are presented in Sec.~\ref{Sec7}, which contains one case of nonlocal form factor and three cases of local higher-derivative gravity with increasing complexity (up to the case of the most general polynomial form factor, including an arbitrary number of complex and multiple roots). Finally, Sec.~\ref{Sec8} contains a summary of the results and our conclusion.
Throughout this paper, we use the same sign conventions of~\cite{BreTibLiv}, and we adopt the unit system such that $c = 1$ and $\hslash  = 1$.


\section{Effective field equations: The general solution for $A(r)$ and $B(r)$}
\label{Sec2}

As mentioned in the Introduction, our discussion of the higher-derivative modifications of the Schwarzschild metric is based on the system of equations~\eq{nlEE2}.
On the left-hand side we have the Einstein tensor, whose nonzero components evaluated for the generic metric~\eq{metric} are
\beq
\n{Gtt}
G^t {}_t = \frac{A'}{r}+\frac{A}{r^2}-\frac{1}{r^2}
,
\eeq
\beq
\n{Grr}
G^r {}_r = G^t {}_t  + \frac{A B'}{r}
,
\eeq
and
\beq
\n{Gang}
\begin{split}
G^\th {}_\th  = G^\phi {}_\phi  = & \, \frac{A'}{r} + \frac{1}{2}\frac{A B'}{r} + \frac{3}{4} A' B' + \frac{1}{4} A B'^2  + \frac{1}{2} A^{\prime\prime} + \frac{1}{2} A B^{\prime\prime}
.
\end{split}
\eeq
Therefore, using the effective energy momentum tensor~\eq{effT}, the field equations~\eq{nlEE2} are equivalent to
\beq
\n{EqGtt}
\frac{A'}{r}+\frac{A}{r^2}-\frac{1}{r^2} = - 8\pi G \rho_{\rm eff}
,
\eeq
\beq
\n{EqGrr}
\frac{A B'}{r} = 8\pi G ( \rho_{\rm eff} + p_r )
\eeq
and
\beq
\n{EqGang}
\frac{A'}{r} + \frac{1}{2}\frac{A B'}{r} + \frac{3}{4} A' B' + \frac{1}{4} A B'^2 
  + \frac{1}{2} A^{\prime\prime} + \frac{1}{2} A B^{\prime\prime} = 8\pi G  p_\th
.
\eeq
The first and the last equations are just the $tt$ and $\th\th$ components of~\eq{nlEE2}, while~\eq{EqGrr} is obtained from the $rr$ equation by subtracting~\eq{Gtt} from \eq{Grr}.

The formal solution for $A(r)$ can be directly obtained from Eq.~\eq{EqGtt}. It only depends on the effective delta source, and it is given by
\beq
\n{asol}
A(r) = 1 - \frac{2G m(r)}{r}
,
\eeq
where the \emph{mass function}
\beq
\n{massf}
m(r) = 4 \pi \int_0^r \rd x \, x^2 \rho_{\rm eff} (x)
\eeq
represents the effective mass inside a sphere of radius $r$.

Given the solution~\eq{asol} for $A(r)$, the function $B(r)$ can be obtained through~\eq{EqGrr}. The formal solution reads 
\beq
\n{Bsol}
B(r) = B_0 + 8\pi G \int_{r_0}^r \rd x \, \frac{x \left[ \rho_{\rm eff} (x) + p_r (x) \right]}{A(x)}
.
\eeq
The constant $B_0 = B(r_0)$ is responsible for a nontrivial time delay between the position $r$ and an observer at $r_0$ (for more details, see the discussion in~\cite{Frolov:2016pav}). Of course, one can always fix $B_0$ by imposing the asymptotic condition $\lim_{r\to\infty} B(r) = 0$, so that the metric reproduces the Minkowski 
geometry at infinity---provided that the effective pressures tend to zero sufficiently fast for large $r$. The constant $B_0$ can also be changed by a redefinition of the time coordinate $t$, and for these reasons hereafter we shall take its value according to the convenience without further elaborations.

Finally, since the solution for the function $B(r)$ depends on the choice of the pressure components of the effective energy-momentum tensor $\tilde{T}^{\mu}{}_{\nu}$, it is mandatory to comment on these other components. 
One of the simplest possible choices is to fix the effective radial pressure $p_r$ via an equation of state involving this component and the effective source $\rho_{\rm eff}$. In this case, both $A(r)$ and $B(r)$ are directly determined in terms of the source $\rho_{\rm eff}$. Then, the component $p_\th$ is given by the last equation of motion~\eq{EqGang}. This is completely equivalent to fixing $p_\th$ using the effective conservation equation $\nabla_\mu \tilde{T}^\mu {}_r = 0$, namely,
\beq
\n{Conserva}
p_r^\prime = - \frac{1}{2} \left( \frac{A^\prime}{A} + B^\prime \right)  \left( p_r + \rho_{\rm eff} \right)  - \frac{2}{r} \left( p_r - p_\th \right),
\eeq
from which one can easily determine $p_\th$ once $\rho_{\rm eff}$ and $p_r$ (and, consequently, $A$ and $B$) are specified. Indeed, if this equation is satisfied, the $tt$ and $rr$ components of~\eq{nlEE2} imply in~\eq{EqGang}. 
In Sec.~\ref{Sec4}, we discuss possible choices of equations of state in the form $p_r = p_r (\rho_{\rm eff})$, their motivations, and the properties of the corresponding function $B(r)$.

As a conclusion of this section, the solution of the effective field equations~\eq{nlEE2} can be expressed as 
\beq
\label{solucao_mais_geral_possivel}
\rd s^2 = - \left(1-\frac{2G m(r)}{r} \right) \exp \left( B_0 + 8\pi G \int_{r_0}^r \rd x \, \frac{x \left[ \rho_{\rm eff} (x) + p_r (x) \right]}{A(x)}  \right) \rd t^2 + {\left(1-\frac{2G m(r)}{r} \right)}^{-1} \rd r^2 + r^2 \rd \Omega^2
.
\eeq


\section{Newtonian limit}
\label{Sec5}

Here we make a brief digression to show how to reproduce the Newtonian-limit metric~\eq{metric-New} using the field equations~\eq{EqGtt}--\eq{EqGang}. 
In the linear approximation we write 
\beq
A(r) \longmapsto 1 +  a(r),
\qquad
B(r) \longmapsto  b(r)
\eeq
and only keep quantities up to the first order in $a(r)$, $b(r)$ and their derivatives. Also, in the nonrelativistic limit, one sets $p_r = p_\th = 0$.  Thus, the linearized Eqs.~\eq{EqGtt} and \eq{EqGrr} are
\begin{align}
&
\frac{a (r)}{r^2}+\frac{a '(r)}{r} = - 8 \pi G \rho_{\rm eff}(r)
,
\n{eqalin}
\\
&
\frac{b '(r)}{r} = 8 \pi G \rho_{\rm eff}(r)
,
\n{eqblin}
\end{align}
whose solutions are the functions entering the linearized version of the metric \eq{metric}, i.e.,
\beq
\n{linme}
\rd s^2 = -[ 1 + a(r) + b(r) ] \rd t^2 + [1 - a(r)] \rd r^2 + r^2 \rd \Omega^2
.
\eeq

Since Eq.~\eq{EqGtt} is already linear in the metric function $A(r)$, there is actually no approximation involved in \eq{eqalin}, which is only a rewriting of the former in terms of the function $a(r) = A(r) - 1$. Its solution is
\beq
\n{alinsol}
a(r) = -\frac{2 G m(r)}{r}
,
\eeq
with $m(r)$ defined in \eq{massf}, while $b(r)$ is easily obtained from~\eq{eqblin},
\beq
\n{blinsol}
b(r) = b_0 +  8 \pi G \int_{r_0}^r \rd x \, x \, \rho_{\rm eff} (x) .
\eeq

Finally, since $\rho_{\rm eff}$, $a(r)$, and $b(r)$  are proportional to $M$, the nontrivial component~\eq{Conserva} of the continuity equation,
\beq
\na_{\mu} \tilde{T}^{\mu} {}_{r} = \frac{1}{2} \left( \frac{A^\prime}{A} + B^\prime \right) \rho_{\rm eff}  = \frac{1}{2}  ( a^\prime + b^\prime )  \rho_{\rm eff},
\eeq
is already $O(M^2)$, implying that it is verified within the approximation. 

The Newtonian potential $\ph(r)$ is defined through the relation
\beq
g_{tt} = - \left[ 1 + 2 \ph(r) \right] ,
\eeq
thus from~\eq{linme} we obtain
\beq
\label{phiab}
\ph(r) = \frac{a(r) + b(r)}{2}
.
\eeq
Taking the derivative of the previous equation, we find
\beq
\ph^\prime(r) = \frac{a'(r) + b'(r)}{2} = -\frac{a(r)}{2r}
,
\eeq
where we used \eq{eqalin} and~\eq{eqblin}. Therefore,
\beq
\label{aebf}
a(r) = - 2 \ph'(r) r \qquad \text{and} \qquad  b(r) = 2 \left[  \ph(r) + \ph'(r) r \right] 
.
\eeq
Substituting these expressions in~\eq{linme}, we obtain the Newtonian metric in the form \eq{metric-New}, presented in the Introduction.


\section{Curvature regularity and suppression of nonlinearities near $r=0$}
\label{singu}
\label{Sec6}

Before discussing the general properties of the solutions for $A(r)$ and $B(r)$ obtained in Sec.~\ref{Sec2}, it is instructive to review what conditions such functions should satisfy to regularize the curvature and curvature-derivative invariants. We also elaborate more on the appealing idea that nonlinearities are suppressed close to $r=0$. We show that under some general assumptions on $A(r)$ and $B(r)$, the behavior of the curvature invariants for sufficiently small $r$ is, indeed, well approximated by the Newtonian-limit solutions.


\subsection{Regularity of curvature invariants}
\label{Sec6.1}

The regularity analysis of the curvature invariants constructed only by contracting an arbitrary number of Riemann and metric tensors can be greatly simplified by noticing that any of such scalars can be 
expressed as a combination of the components ${R^{\mu\nu}}_{\al\beta}$ (i.e., as a sum of products of these components only, without the need of using the metric tensor)~\cite{Bronnikov:2012wsj}. 
For a static spherically symmetric metric, ${R^{\mu\nu}}_{\al\beta}$ has only four independent components, which in the case of a metric in the form~\eq{metric} are given by 
\begin{subequations} \label{CiDef}
\begin{align}
\label{C1gen}
K_1 & \equiv  {R^{tr}}_{tr} = -\frac{A''}{2}-\frac{1}{2} A B'' -\frac{3}{4} A' B'-\frac{1}{4} A B'^2  ,
\\
\label{C2gen}
K_2 & \equiv  {R^{t\th}}_{t\th} = {R^{t\phi}}_{t\phi}  = -\frac{A'+A B'}{2 r} ,
\\
\label{C3gen}
K_3 & \equiv  {R^{r\th}}_{r\th} = {R^{r\phi}}_{r\phi} =  -\frac{A'}{2 r} ,
\\
\label{C4gen}
K_4 & \equiv  {R^{\th\phi}}_{\th\phi} = -\frac{A-1}{r^2} .
\end{align}
\end{subequations}
If the four functions $K_i(r)$ are regular, all the curvature invariants without covariant derivatives are bounded as well.

In an equivalent way, one can use the Kretschmann scalar, 
\beq
\n{Kre}
K \equiv R_{\mu\nu\al\beta} R^{\mu\nu\al\beta} = 4 K_1^2 + 8 K_2^2 + 8 K_3^2 + 4 K_4^2 ,
\eeq
to investigate the regularity of a whole set of curvature invariants. Indeed, since it is the sum of squares, for its regularity it is necessary and sufficient that all functions $K_i (r)$ are finite. Thus, if $K$ is bounded, all the $K_i (r)$ are bounded as well and, as a consequence, so is any scalar formed only by curvature contractions~\cite{Bronnikov:2012wsj} (e.g., the scalar curvature $R$, the square of the Ricci tensor $R_{\mu\nu} R^{\mu\nu}$, the square of the Weyl tensor $C_{\mu\nu\al\beta} C^{\mu\nu\al\beta}$ or any polynomial scalar constructed with them). However, from the practical point of view, in order to verify the regularity of a given metric, it is simpler to work with the short expressions in~\eq{CiDef} rather than the one in~\eq{Kre}.

To analyze the behavior for small $r$, let us assume that $A(r)$ and $B(r)$ are analytic\footnote{This is true for the solution~\eq{solucao_mais_geral_possivel} in the models this paper concerns, as shown in Secs.~\ref{Sec3} and~\ref{Sec4} below. For a discussion of the regularity of invariants with non-analytic metrics see,~e.g.,~\cite{Giacchini:2021pmr}.}
around $r = 0$ and apply the power series expansion,
\beq
\n{ps_geral}
A(r) = \sum_{n = 0}^\infty a_n r^n,
\qquad
B(r) = \sum_{n = 0}^\infty b_n r^n,
\eeq
into the expressions for the functions $K_i(r)$:
\begin{subequations}
\begin{align}
K_1 & =  -  a_2 - a_0 b_2 - \frac{b_1 (3 a_1 + a_0 b_1 )}{4} + O(r) ,
\\
K_2 & =  - \frac{a_1 + a_0   b_1 }{2 r} -  a_2 - \frac{ a_1 b_1 }{2} - a_0   b_2 + O(r) ,
\\
K_3 & =   - \frac{ a_1 }{2 r} - a_2  + O(r) ,
\\
K_4 & =   - \frac{ 1 - a_0 }{r^2} - \frac{ a_1 }{r} - a_2  + O(r) .
\end{align}
\end{subequations}
Therefore, the regularity of the curvature invariants requires 
\beq
\n{finito}
a_0 = 1 \qquad \mbox{and} \qquad a_1 = b_1 = 0,
\eeq
or, in other words,
\begin{align}
\n{A(0)=1}
A(0) &= 1, \\ 
\n{AB'(0)=0}
A'(0) &= B'(0) = 0.
\end{align}
In the next sections we show that these conditions are fulfilled for the general solution~\eq{solucao_mais_geral_possivel} in any theory whose form factor is such that $f(-k^2) \sim k^4$ or faster for large $k$---for example, for effective sources of models that contain six or more metric derivatives.

To close this section, we briefly comment on the regularity of curvature-derivative invariants. As highlighted in Refs.~\cite{Nos6der,Giacchini:2021pmr}, scalars that are polynomial in curvature tensors and their covariant derivatives can display a singular behavior at $r=0$ if the regular metric of the form~\eqref{metric} contains odd powers of $r$ in the Taylor expansion~\eq{ps_geral} of its components. Thus, some of the considerations can be stated in a simple form by using the definition of order of regularity of a function, as introduced in~\cite{Nos6der}:

\vskip 2mm 
\noindent
\textbf{Definition 1.} Given a continuous bounded function $\xi:[0,+\infty)\to\mathbb{R}$, we say that 
$\xi(r)$ is \emph{$p$-regular} for an integer $p\geqslant 1$ if 
$\xi(r)$ is at least $2p$-times continuously differentiable and the first $p$ odd-order derivatives of $\xi(r)$ vanish as $r \to 0$, namely,
\beq
0 \leqslant n \leqslant p-1 \quad \Longrightarrow \quad \lim_{r \to 0} \, \xi^{(2n+1)}(r) = 0 .
\eeq
If these conditions are satisfied, we shall also say that $p$ is the \emph{order of regularity} of $\xi$. This extends the definition of a regular, continuous function (which can also be called 0-regular) in a way that is convenient for studying the regularity of the spacetime.
\vskip 2mm 

Indeed, if a function $\xi:[0,+\infty)\to\mathbb{R}$ is $p$-regular, this means that also the map $\Xi:\mathbb{R}^3\to\mathbb{R}$ given by $\Xi(\vec{r}) = \xi(\vert \vec{r} \vert)$ is of class $C^{2p}$. On the other hand, if $\lim_{r \to 0} \, \xi^{(2p+1)}(r) \neq 0$, then (switching to Cartesian coordinates) we have
\beq
\lim_{x \to 0} \partial^{2p+1}_x \Xi (x,0,0) = \pm \xi^{(2p+1)}(0),
\eeq
depending on whether $x=0$ is approached from the left or the right. This means that $\Xi$ is not $(2p+1)$-times differentiable, and if we take one of the metric functions $A(r)$ or $B(r)$ for $\xi$, this implies that the metric is not smooth at $r=0$.

General evidences of a relation between the order of the first odd power of $r$ in the series and the minimal number of covariant derivatives in a divergent curvature-derivative scalar were presented in the works~\cite{Nos6der,Giacchini:2021pmr}. (See~\cite{Nos6der} for a complete treatment of the problem at linear level, and~\cite{Giacchini:2021pmr} for the detailed consideration of invariants of the type $\Box^n R$ at nonlinear level and further examples.) 
Taking into account these results, if the first odd power of $r$ in the series expansion of the functions $A(r)$ and $B(r)$ is at order $r^{2n+1}$ for some $n\in\mathbb{N}$, then there might exist curvature-derivative scalars with $2n$ covariant derivatives of curvature tensors that diverge at $r=0$.
All local curvature-derivative scalars are expected to display a nonsingular behavior only if $A(r)$ and $B(r)$ are $\infty$-regular, i.e., if the metric components are {\it even functions}.

A simple useful example is provided by the scalar $\Box R$, which for the metric~\eq{metric} reads
\beq
\n{boquisErre}
\begin{split}
\Box R \, = \, &  \frac{1}{4 r^4} \,
\bigg( 16 A 
- 16 A^2 
- 16 r A'
+ 32 r A A'
+ 8 r^2 A'^2
- 8 r A B'
+ 8 r A^2 B' 
+ 12 r^2 A A' B' 
\\
& 
-8 r^3 A'^2 B' 
+4 r^2 A^2 B'^2 
-8 r^3 A A' B'^2
-2 r^4 A'^2 B'^2 
-r^4 A A' B'^3
-8 r^2 A A''
-16 r^3 A' A''
\\
& 
-28 r^3 A B' A'' 
- 6 r^4 A' B' A'' 
-5 r^4 A B'^2 A''
-44 r^3 A A' B''
-10 r^4 A'^2 B''
-12 r^3 A^2 B' B'' 
\\
& 
-17 r^4 A A' B' B'' 
-2 r^4 A^2 B'^2 B''
-16 r^4 A A'' B'' 
-4 r^4 A^2 B''^2 
-24 r^3 A A^{(3)}
-4 r^4 A' A^{(3)}
\\
& 
-8 r^4 A B' A^{(3)} 
-16 r^3 A^2 B^{(3)} 
-18 r^4 A A' B^{(3)} 
-6 r^4 A^2 B' B^{(3)}
-4 r^4 A A^{(4)}
-4 r^4 A^2 B^{(4)} \bigg) 
\,.
\end{split}
\eeq
Once more,
to investigate the behavior of the invariant near $r=0$ we use the power series representation~\eq{ps_geral}, already assuming $a_0=1$ and $a_1 = b_1 = 0$. The result is
\beq
\Box R = -\frac{8 (5 a_3 + 3 b_3)}{r} +  O(r^0).
\eeq
Hence, the scalar $\Box R$ diverges like $1/r$ unless $5 a_3 + 3 b_3 = 0$; in particular, it is regular if $a_3 = b_3 = 0$---in other words, if $A(r)$ and $B(r)$ are 2-regular, as noted in~\cite{Nos6der,Giacchini:2021pmr}. 
In the following sections, we show that there exist equations of state for the effective pressures such that this condition is fulfilled in any theory whose form factor is such that $f(-k^2) \sim k^6$ or faster for large $k$ [see, e.g., Eqs.~\eq{AprideG} and~\eq{Bde} below], i.e., for smeared delta sources of models with eight or more derivatives. 
We remark, however, that the regularity of curvature-derivative scalars, which depend on higher-order derivatives of the metric, can be very sensitive to the choice of equation of state; an example of this is presented in Appendix~\ref{App1}.


\subsection{Suppression of nonlinearities}

As stated in the Introduction, some authors have argued that close to $r = 0$, the gravitational field may be very weak, so the linearized solution reproduces the fate of the singularities.
Now, we elaborate more on this idea, showing that this can be the case under some assumptions on the form of the metric components.

First, let us suppose that the spacetime is asymptotically flat such that we can write
\beq
\n{A=1+a}
A(r) = 1 + a(r),
\eeq
with $\lim_{r \to \infty} a(r) = 0$. We also assume that $a(r)$, $B(r)$, and their derivatives are proportional to a parameter such that the term ``linearization'' is understood as the evaluation of quantities at leading order in these functions, since they are already linear in the parameter (i.e., there is no linearization {\it inside} these functions). In the explicit examples we consider in Sec.~\ref{Sec7}, this parameter can be taken, e.g., as the mass\footnote{Note that this criterion is not satisfied, e.g., for the Hayward metric~\cite{Hayward}, because it is not linear in $M$, namely, 
$$
a(r) = - \frac{2M r^2}{r^3 + M^2 L}, \quad B(r) = 0. 
$$
}  
$M$.

The next step is to apply this procedure and calculate the linearized version of the curvature tensor, $({R^{\mu\nu}}_{\al\beta})_{\rm lin}$. Using~\eq{CiDef}, its components read
\begin{subequations}
\begin{align}
K_1^{\rm lin} & =  - \frac{a'' + B''}{2}  ,
\\
K_2^{\rm lin} & =  - \frac{a' + B'}{2 r} ,
\\
K_3^{\rm lin} & =  K_3 ,
\\
K_4^{\rm lin} & =   K_4.
\end{align}
\end{subequations}
Notice that since $K_3$ and $K_4$ are already linear in $A(r)$ and do not depend on $B(r)$, their linearized expressions coincide with the original ones.

The nonlinear part of ${R^{\mu\nu}}_{\al\beta}$ is then proportional to the differences 
\beq
K_i^{\rm non-lin} \equiv K_i - K_i^{\rm lin},
\eeq 
which turn out to be proportional to the derivatives of $B(r)$:
\begin{subequations} \n{CiNonLin}
\begin{align}
K_1^{\rm non-lin}  & =    \frac{1}{4} \left[ 3 a' B' + (1 + a ) B'^2 + 2 a  B'' \right] ,
\\
K_2^{\rm non-lin}  & =    \frac{a  B' }{2 r} ,
\\
K_3^{\rm non-lin} & =  K_4^{\rm non-lin} = 0 .
\end{align}
\end{subequations}
In the same spirit as before, using the power series representation~\eq{ps_geral} for $A(r)$ and $B(r)$ we find, around $r=0$,
\begin{subequations}
\begin{align}
K_1^{\rm non-lin}  & =    \frac{1}{4}  b_1  (3  a_1  + a_0  b_1 ) + ( a_0  -1 )  b_2 + O(r)  ,
\\
K_2^{\rm non-lin}  & =    \frac{(a_0  -1)  b_1 }{2 r} + \frac{ a_1   b_1 }{2} + ( a_0  -1 )  b_2   + O(r) .
\end{align}
\end{subequations}
Therefore, if the curvature invariants are regular, i.e., if the relations in~\eq{finito} are true, we get
\beq
K_i^{\rm non-lin} = O(r),
\eeq 
showing that for small $r$ the nonlinearities in ${R^{\mu\nu}}_{\al\beta}$ are suppressed. Since all the invariants polynomial in curvatures and metric tensors can be build using combinations of the functions $K_i$, it follows that, for small $r$, such scalars behave approximately like the linearized ones. 

Furthermore, if the functions $a(r)$ and $B(r)$ approach the Newtonian solution for small $r$, not only are the nonlinearities suppressed, but the value of the curvature invariants evaluated at $r=0$ is equal to the Newtonian one. This happens, indeed, with the general solution~\eq{solucao_mais_geral_possivel} if the effective pressures tend to zero sufficiently fast for small $r$.
As shown in Sec.~\ref{Sec5}, the solution $a(r)$ of the Newtonian equation of motion~\eq{eqalin} is the same one entering $A(r) = 1 + a(r)$ that solves the nonlinear equations of motion~\eq{nlEE2}; while for $B(r)$, given that $A(0) = 1$, it is always possible to find a small enough $r$ such that $1/A(r) \approx 1$ and, as a result, the solution~\eq{Bsol} approaches~\eq{blinsol} if $p_r$ vanishes sufficiently fast as $r \to 0$. 

It is also worthwhile to note that if $B(r) = 0$, then $K_i^{\rm non-lin} = 0$ always, {\it even for singular metrics}.
This explains why in general relativity the Kretschmann invariant evaluated using either the Schwarzschild solution or the Newtonian-limit metric gives the same result 
\beq 
\label{KSch}
K_{\text{Sch.}} = \frac{48 G^2 M^2}{r^6} .
\eeq

In what concerns the curvature-derivative invariants, the situation is slightly different. Using once more $\Box R$ as an example and proceeding in the same way as before, we split~\eq{boquisErre} in terms of the linear and nonlinear parts,
\beq
(\Box R)_{\rm lin} = - \frac{4 a}{r^4} + \frac{4 a'}{r^3} - \frac{2 a''}{r^2} - \frac{6 a^{(3)} + 4 B^{(3)}}{r}  - a^{(4)} -  B^{(4)}
,
\eeq
\beq
\label{420}
\begin{split}
(\Box R)_{\rm non-lin} \, \equiv \, & \,
\Box R - (\Box R)_{\rm lin}
\\
= \, & \,
a \left( 
-\frac{4 a}{r^4} + \frac{8 a'}{r^3} - \frac{2 a''}{r^2} - \frac{6 a^{(3)}}{r} - a^{(4)} 
\right) 
+ a' \left( 
\frac{2 a'}{r^2} - \frac{4 a''}{r} - a^{(3)} \right) 
- B' \bigg[ 
 \frac{2 a'^2}{r}  + \frac{1}{2} a'^2 B'
\\
&
+ \frac{3}{2} a' a'' - (1+a) \bigg( \frac{2  a}{r^3}  + \frac{(1 + a) B' + 3  a'}{r^2}  - \frac{2  a' B' - 7  a''}{r} 
- \frac{1}{4}  a' B'^2 - \frac{5}{4}  B' a'' - 2  a^{(3)} \bigg) \bigg]
\\
&
- B'' \bigg[  \frac{5}{2} a'^2 + (1+a) \bigg( \frac{ 11 a' + 3 (1+a) B'}{r} + 4  a''   + \frac{17}{4} a' B'  + \frac{1}{2} (1+a) ( B'^2  + 2 B'') \bigg) \bigg]
\\
&
- B^{(3)} \bigg[ \frac{4 a (2+a)}{r} + \frac{9}{2} a a' + \frac{3}{2} (1+a)^2 B' \bigg] - B^{(4)} a (2+a)  ,
\end{split}
\eeq
and using the series representation~\eq{ps_geral} for regular metrics (i.e., with $a_0=1$, $a_1=b_1=0$) we get
\begin{align}
\n{LLpfc}
&
(\Box R)_{\rm lin} = -\frac{8 (5 a_3 + 3 b_3)}{r}  - 60 (3 a_4 +2 b_4) +  O(r)
,
\\
&
\n{nlpfc}
(\Box R)_{\rm non-lin} = -12  b_2 \left(6 a_2+b_2\right) + O(r).
\end{align}
Therefore, if $\Box R$ is finite, that is it, if $5 a_3 + 3 b_3 = 0$, we see that its linear and nonlinear parts approach a constant when $r \to 0$. The behavior of $(\Box R)_{\rm non-lin}$ marks a difference between curvature-derivative and curvature invariants since the nonlinear part of the latter goes to zero when $r \to 0$. Note, however, that if $(\Box R)_{\rm lin}$ is finite at $r=0$, the entire $\Box R$ is also finite (because its nonlinear part cannot generate a singularity).

It is also interesting to notice that, because of the covariant derivatives, $(\Box R)_{\rm non-lin}$ contains terms that do not depend on the function $B(r)$ [see Eq.~\eqref{420}]. This is different from what happens in the case of scalars without covariant derivatives. Therefore, even if $B(r) = 0$, $\Box R$ can receive a contribution from the nonlinear part of the metric.


\section{General properties of $A(r)$}
\label{Sec.RA}
\label{Sec3}

In this section, we discuss some properties of \eq{asol} at a general level, without particularizing to any gravitational model effective source. As discussed in the Introduction, by just considering the asymptotic behavior of the action form factor~$F(z)$ [or, equivalently, $f(z)$ in~\eq{fzinho}], we can explain many important physical properties of the solutions that hold true for large classes of higher-derivative effective sources. For instance, we are particularly interested in the absence (or the presence) of an event horizon, of singularities in the curvature invariants,  as well as in the behavior of the solution for large $r$. All the results presented in general form in this section can be seen realized in concrete examples in Sec.~\ref{Sec7}, where we choose some representative form factors and obtain the explicit expressions.

To start, let us recall some relevant results on the smeared delta source $\rho_{\rm eff}(r)$; afterward, we analyze the consequences of these properties to the mass function~\eq{massf} and, finally, to $A(r)$. 
The properties of the effective sources in higher-derivative gravity theories have been intensively studied in the works~\cite{BreTib2,Nos6der} (see also~\cite{BreTibLiv} for an introduction and review). For this reason, we do not repeat their derivation here; we summarize the results as follows.
\begin{itemize}
\item[(1a)] If $f(-k^2) \to 1$ when $k^2 \to 0$ (in other words, if general relativity is recovered in the IR limit), then
\beq
\n{rho-inf}
\lim_{r \to \infty} \rho_{\rm eff}(r) = 0. 
\eeq
In particular, this is related to the solution being asymptotically flat. 
\end{itemize}
Under the assumption (1a) one can also prove that:
\begin{itemize}
\item[(2a)] If there exists a $k_0$ such that $f(-k^2)$ grows as $k^2$ for $k > k_0$, then $\rho_{\rm eff}(r)$ is integrable but it is not bounded. In fact, it diverges at the origin,
\beq
\n{limitefonte4der}
\rho_{\rm eff}(r) \underset{r \to 0}{\sim} \frac{1}{r}
.
\eeq
This case corresponds to the smeared delta sources for Stelle gravity and nonlocal theories that behave as the fourth-derivative theory in the UV. This type of source was also considered in the study of black holes in models with generalized uncertainty principle~\cite{Knipfer:2019pgi}. 

\item[(3a)] On the other hand, if $f(-k^2)$ grows at least as fast as $k^4$ for sufficiently large $k$, the effective source is integrable and finite everywhere, i.e., $\rho_{\rm eff} (r)$ is at least 0-regular. 
In this situation, $\rho_{\rm eff}(r)$ has an absolute maximum at $r=0$,
\beq
\n{rho-r0}
{\rm max} \left\lbrace \rho_{\rm eff} (r)\right\rbrace   = \lim_{r \to 0} \rho_{\rm eff}(r)
.
\eeq
This is the case of local models with six or more metric derivatives and most nonlocal models (namely, those whose propagator is suppressed in the UV at least as fast as in the sixth-derivative gravity).

\item[(4a)] If the function $f(-k^2)$ asymptotically grows at least as fast as $k^{4+2N}$ for an integer
$N \geqslant 1$, then the effective source $\rho_{\rm eff}(r)$ is (at least) $N$-regular (see Definition 1 in Sec.~\ref{Sec6.1}). In other words, it is $2N$ times
differentiable and their first $N$ odd-order derivatives vanish at the origin,
\beq
\frac{\rd^{2n+1} }{\rd r^{2n+1}} \, \rho_{\rm eff} (r) \Big|_{r=0}  = 0 
\qquad \text{for} \qquad n = 0,\cdots, N-1
.
\eeq 

\item[(5a)] As a corollary of the previous statement, if $f(-k^2)$ asymptotically grows faster than any polynomial, then $\rho_{\rm eff}(r)$ is $\infty$-regular, i.e., it is an even function of $r$. This happens, e.g., in the nonlocal models with exponential form factors,  $f(-k^2) = e^{(k^2/\mu^2)^\ell}$ for $\ell\in\mathbb{N}$.
\end{itemize}


From the above results it follows that, for any higher-derivative model form factor, the mass function satisfies~\cite{BreTib2}
\beq
\n{limM1}
\lim_{r\to\infty} m(r) = M.
\eeq
Therefore, for a distant observer, the solution~\eq{asol} for $A(r)$ reduces to Schwarzschild\footnote{In Sec.~\ref{Sec4} we show that, under these circumstances, $B(r)$ tends to a constant; hence, not only $A(r)$, but the geometry itself actually reduces to Schwarzschild.} with mass $M$. On the other hand, for small $r$, the limits~\eq{limitefonte4der} and \eq{rho-r0} together with the definition~\eq{massf} imply that
 
\beq
\n{limM2} 
m(r) \underset{r \to 0}{\sim }  \left\{\begin{array}{rl}
r^2, \,\, &\mbox{if }  f(-k^2) \sim k^2 \,\, \text{for large\,} k \,,
\\
r^3, \,\, &\mbox{if }  f(-k^2) \sim k^4 \, (\mbox{or faster) for large}\, k \,.
\end{array}\right.
\eeq
As consequence, combining \eq{limM1} and \eq{limM2} with \eq{asol} we get
\beq
\n{ALim}
\lim_{r \to 0} A(r) = \lim_{r \to \infty} A(r) = 1.
\eeq 
Notice that this proves the regularity condition of Eq.~\eq{A(0)=1}.
Being a continuously differentiable function, and given \eq{ALim}, it follows that $A(r)$ is bounded for any higher-derivative model effective source. Thus, there is a certain critical mass $M_\text{c}$ such that $M < M_\text{c}$ implies $A(r) > 0$ for all $r$.  

The existence of $M_\text{c}$ means that there is a \textit{mass gap} for the solution to be a black hole. 
Indeed, the presence of an event horizon is related to the invariant
\beq
\n{nar}
(\na r)^2 = A(r) ,
\eeq 
that only  depends on the function $A(r)$.  
The event horizon is given by the largest root of $A(r) = 0$, and the metric does not describe a black hole if $A(r)$ is strictly positive. 

A direct way of verifying the existence of the critical mass $M_\text{c}$ for effective sources is to consider the unitary mass function, defined by
\beq
\tilde{m}(r) = \frac{m(r)}{M},
\eeq
\emph{which does not depend on} $M$ [see Eq.~\eq{effsource}] and it is such that $\lim_{r \to \infty} \tilde{m}(r) = 1$.
Notice that the function $\tilde{m}(r)/r$ must assume a maximum value at some point $r_{*}$ (since it tends to zero in the limits of large and small $r$, see Eq.~\eq{ALim}, and there is a region where it is positive). 
Thus, if $M$ is such that 
\beq
2GM \frac{\tilde{m}(r_{*})}{r_{*}} 
< 1,
\eeq 
the function $A(r)$ is strictly positive, and the metric does not have horizons. The critical mass is, therefore,
\beq
\n{Mcritial-gen}
M_\text{c} = \frac{r_{*}}{2G \tilde{m}(r_{*})}.
\eeq
To summarize, if $f(-k^2)$ grows at least as fast as $k^2$ for sufficiently large $k$, that is it, {\it for smeared delta sources of any higher-derivative gravity theory}, there exists a mass gap for a solution describing a black hole.

In what concerns the solutions that do describe black holes (i.e., if $M > M_{\rm c}$), we notice that, because of~\eq{ALim}, the function $A(r)$ changes sign an even number of times. This means that the solution also contains at least one inner horizon besides the event horizon. The exact number of horizons depends on the function $f(-k^2)$ and on the value of $M$, but it is always an even number. In the limiting case $M = M_{\rm c}$, a pair of horizons merge into one extremal horizon.

In addition, it is worth noticing that if $\tilde{m}(r) \leqslant 1$ for all $r$, then  $m(r) \leqslant M$ and the position of the event horizon $r_+$ is bounded by the Schwarzschild radius,
\beq
\n{bound}
r_+ \leqslant r_{\rm s} = 2GM.
\eeq
In general, if $\rho_{\rm{eff}}(r)$ is positive, the inequality~\eq{bound} is always true. This statement can be proved by observing that
if $\rho_{\rm{eff}}(r)$ is positive, then the mass function $m(r)$ is monotone. In fact, 
\beq
\label{implication}
\rho_{\rm{eff}}(r) \geqslant 0 \qquad \Longrightarrow \qquad m^\prime(r) = 4\pi r^2 \rho_{\rm{eff}}(r) \geqslant 0 ,
\eeq
whence $m(r)$ is a monotonically increasing function. Then, from Eq.~\eq{limM1} it follows $m(r) \leqslant M$. On the other hand, if there is a region where $\rho_{\rm{eff}}(r) < 0$, then $m(r)$ must have oscillations, and it may assume values larger than $M$ (see~\cite{Modesto12,Zhang14,Bambi:2016wmo,NosLW} for explicit examples).

The value of the critical mass $M_{\rm c}$ is related to the massive parameters of the model (see~ Sec.~\ref{Sec7} for explicit examples). If $M \gg M_{\rm c}$ and those parameters are such that the smearing of the source is negligible at the scale $2GM$, then we expect the solution to be very close to the Schwarzschild solution at the outer horizon. In particular, in the context of Eq.~\eqref{bound}, $r_+$ approaches $r_{\rm s}$ from below.

It is important to keep in mind that the linearity of the function $A(r)$ in the mass parameter $M$ plays a central role when deriving the above conclusions regarding the number of horizons, oscillations of the effective mass function, and the existence of a black hole mass gap. This underlying assumption can be traced back to Eq.~\eqref{EqGtt} and the form of the effective source~\eqref{effsource}. An intriguing question is whether such features of the effective solutions are also present in the solutions of the complete field equations of the original higher-derivative/nonlocal gravity. The results on exact solutions of sixth- and higher-derivative gravities so far available in the literature are not enough to clarify the situation, for they mainly concern either the weak-field regime outside a horizon or local aspects of the solutions~\cite{Holdom:2002xy,Pawlowski:2023dda,Daas:2024pxs,Giacchini:2024exc}. Nevertheless, we remark that even for general functions $A(r)$, asymptotically flat, everywhere regular geometries must have an even number of horizons~\cite{Holdom:2002xy}.

To close this section on general features of the function $A(r)$, we present some of the properties of its derivatives, which  are related to the occurrence of singularities in the curvature invariants at $r=0$, as discussed in Sec.~\ref{Sec6}.

Applying Taylor's theorem and taking into account the results (3a) and (4a) on the behavior of the effective delta source near $r=0$, it follows that: If the function $f(-k^2)$ asymptotically grows at least as fast as $k^{4+2N}$ for an integer
$N \geqslant 0$, then $A(r)$ is at least $(N+1)$-regular, i.e., it is $2N+2$ times
differentiable and 
\beq
\n{AprideG}
\frac{\rd^{2n+1} }{\rd r^{2n+1}} \, A(r) \Big|_{r=0}  = 0 
\qquad \text{for} \qquad n = 0,\ldots, N
.
\eeq
Notice that this result also covers the cases of the nonlocal form factors by Kuz'min and Tomboulis~\cite{Kuzmin,Tomboulis}, which behave like polynomials in the UV.
In particular, if $f(-k^2)$ asymptotically grows at least as fast as $k^{4}$ (like in any polynomial model), we get $A'(0) = 0$, which is a necessary condition for the absence of curvature singularities at $r=0$ [see the discussion involving Eq.~\ref{AB'(0)=0}].

Finally, as a corollary, if $f(-k^2)$ asymptotically grows faster than any polynomial, $A(r)$ is an even function---which also implies that the effective mass $m(r)$ is an odd function. This can be proved, alternatively, from the result (5a) on the effective delta source:
\beq
\n{m-odd}
\rho_{\rm eff}(-r) = \rho_{\rm eff}(r) \qquad \Longrightarrow \qquad m(-r) = - 4\pi \int_{0}^r \rd x \, x^2 \rho_{\rm eff} (-x) = - m(r),
\eeq
where we made the change of variable $x \mapsto -x$ in the integration; thus, $A(-r) = A(r)$. As recently discussed in~\cite{Nos6der,Giacchini:2021pmr}, if $A(r)$ and $B(r)$ are even analytic functions, not only the curvature invariants are finite at $r=0$, but also all the curvature-derivative invariants. The analogous result for $B(r)$ is going to be derived in the next section.


\section{General properties of $B(r)$}
\label{Sec4}

In this section, we analyze several choices for fixing the effective pressures and underlying motivations. For each, we present the general properties of the function $B(r)$ associated with it.


\subsection{Case of null pressures} 
\label{nopressure}
\label{Sec4.1}

Considering
\beq
p_r = p_\th = 0,
\eeq
the solution~\eq{Bsol} for $B(r)$ becomes
\beq
\n{bsol2}
B(r) = B_0 +  8 \pi G \int_{r_0}^r \rd x \, \frac{x \, \rho_{\rm eff} (x)}{A(x)} .
\eeq

The properties of the function $B(r)$ defined in~\eq{bsol2} can be directly deduced from the results concerning the effective source (mentioned in Sec.~\ref{Sec.RA} and proved in~\cite{BreTib2,Nos6der}) and the results obtained for $A(r)$ in the previous section. Namely, if $A(r) > 0\,$ $\forall \, r$ then $B(r)$ is bounded for any higher-derivative model. Moreover, if $f(-k^2) \sim k^2$ for large $k$, then $B(r)$ is differentiable but it is not 1-regular, as
$
B'(0) \neq 0;
$
while if $f(-k^2) $ grows at least as fast as $k^{4+2N}$ for a certain $N \geqslant 0$ and sufficiently large $k$, $B(r)$ is at least $(N+1)$-regular, hence,
\beq
\n{Bde}
\frac{\rd^{2n+1} }{\rd r^{2n+1}} \, B(r) \Big|_{r=0}  = 0 
\qquad \text{for} \qquad n = 0,\cdots, N
.
\eeq
In particular, for any model with six or more derivatives, i.e., if $f(-k^2) \sim k^4$ or faster for large $k$,
$
B'(0) = 0.
$ 
Together with~\eq{AprideG}, this last result is responsible for canceling the curvature singularity at $r=0$ for models with $f(-k^2)$ growing at least as fast as $k^4$ [see the discussion involving Eq.~\ref{AB'(0)=0}]. Moreover, if $f(-k^2)$ grows faster than any polynomial, there exists a neighborhood of $r=0$ where $B(r)$ is an even function. This can be proved by setting $r_0 = 0$ in~\eq{bsol2}, making the change of integration variable $x \mapsto -x$, and using that for this type of form factor $\rho_{\rm eff}(r)$ and $A(r)$ are also even functions.

The solution~\eq{asol} together with~\eq{bsol2} corresponds to the case where Eq.~\eq{nlEE2} serves as a ``small-curvature approximation'' of the full nonlinear equations of motion for the theory~\eq{action} sourced by a pointlike mass,
\beq
\n{Tori}
T_{\mu\nu}(\vec{r}) =  M \de^t_\mu \de^t_\nu \de^{(3)}(\vec{r}).
\eeq
Indeed, if the original source is a pointlike particle, the inversion of the operator $f(z)$ in \eq{cu}, ignoring the terms $O(R^2_{\cdots})$ of higher order in curvatures, automatically gives the effective energy-momentum tensor~\eq{effT} with $p_r = p_\th = 0$. 
However, a careful reader will notice that in this situation the effective energy-momentum tensor $\tilde{T}^\mu {}_\nu$ is not conserved [or, equivalently, the equation $G^\th {}_{\th} = 0$ is not satisfied, see~\eq{EqGang}]. In other words, this choice of effective pressures makes the system \eq{EqGtt}-\eq{EqGang} over-determined. Approximate solutions can occur, nevertheless: 
The conservation equation~\eq{Conserva} might hold only \emph{under the approximation}, namely, in regions where the spacetime curvature is small---that is, in the whole spacetime (if $M < M_c$) or in the limits of short and large distances (if $M > M_c$), as we discuss in what follows.

Firstly, for $M < M_c$, the situation is similar to what happens in the Newtonian approximation (see the discussion in Sec.~\ref{Sec5}) since the conservation equation~\eq{Conserva} for $p_r = p_\th=0$ reduces to 
\beq
-\frac12 \left(\frac{A'}{A} + B' \right) \rho_{\rm eff} = O (M^2).
\eeq
Similarly, using~\eq{asol} and~\eq{bsol2} in~\eq{Gang}, it is possible to prove (after some manipulations) that
\beq
G^\th {}_\th = O (M^2),
\eeq
which means that the field equations are satisfied in a good approximation in the whole spacetime if the mass $M$ is sufficiently small. How small the mass needs to be depends on the critical mass $M_c$, because for $M < M_c$ the functions $A(r)$ and $B(r)$ are bounded, and so are the curvatures. Indeed, in this case, the spacetime curvature is limited by a certain value and might not grow too much. This will become more evident in the specific examples of Sec.~\ref{Sec7}.

On the other hand, for $M > M_c$ the solution~\eq{bsol2} is ill-defined at the position of the horizons, given the occurrence of the term $1/A(r)$ in the integrand. The dominance of the singular structure $1/A(r)$ in~\eq{bsol2} characterizes the high-curvature regions of the spacetime, where the ``small-curvature approximation'' breaks down. Hence, the expression~\eq{Bsol} can be used as long as the interval $[r_0,r]$ does not cross any horizon. 
Assuming that $A(r)$ and $B(r)$ are analytic functions,\footnote{As stated in the Introduction, in this work we mainly focus on cases where the form factor $f(-k^2)$ is an analytic function that grows at least as fast as $k^4$ for large $k$. In this context, from the above results, $A(r)$ and $B(r)$ are analytic indeed.} far from the horizons and near $r=0$
we can use their power series representation~\eq{ps_geral}, which substituted into~\eq{Gang} give
\beq
G^\th {}_\th = \frac{1-a_0}{r^2}
- \frac{a_0 b_1 + 2 a_1}{2r}
+\frac{1}{4} \left( a_0 b_1^2 + a_1 b_1 \right)
+ O (r)
.
\eeq
But, from Eqs.~\eq{ALim},~\eq{AprideG} and~\eq{Bde} we have
$a_0 = 1$ and $a_1 = b_1 = 0$, such that 
\beq
G^\th {}_\th = O(r).
\eeq 
This means that the effective field equations are satisfied in a good approximation for a sufficiently small $r$ around the point $r=0$ for any value of the mass $M$. Note that this is true only for the effective sources of models with \emph{more} than four derivatives of the metric in the action. In fourth-derivative gravity, we have $a_1, b_1 \neq 0$; therefore, the effective equations~\eq{nlEE2} will not reproduce the properties described in~\cite{Lu:2015cqa,Stelle15PRD} of Stelle gravity for small $r$. Finally, the angular equations are also approximately satisfied for large $r$ because $\lim_{r \to \infty} \rho_{\rm eff} = 0$ implies $\lim_{r \to \infty} G^\th {}_\th = 0$ through the functions $A(r)$ and $B(r)$.


\subsection{Case $p_r = \alpha \, \rho_{\rm eff}$}
\label{Sec-Alpha}
\label{Sec4.2}

If one sets
\beq 
\n{Pr_prop_rho}
p_r = \alpha \, \rho_{\rm eff}
\eeq
for a certain constant $\al$, but leaves the tangential pressure to be determined by the conservation equation \eq{Conserva}, the result for $B(r)$ is
\beq
\n{bsolalpha}
B(r) = B_0 + (1 + \al) 8 \pi G \int_{r_0}^r \rd x \, \frac{x \, \rho_{\rm eff} (x)}{A(x)},
\eeq
while $p_\th$ is given by
\beq
\n{Pth1}
p_\th  =  \frac{1+\al}{4} \left( \frac{A^\prime}{A} + B^\prime \right) r \rho_{\rm eff} + \frac{\alpha}{2} \left( 2 \rho_{\rm eff} + r \rho_{\rm eff}^\prime \right) .
\eeq
Therefore, the functions $A(r)$ and $B(r)$ can be obtained from the equations \eq{EqGtt} and \eq{EqGrr} alone, and the quantity $p_\th$ is constructed afterward. (It is straightforward to verify that, with $p_\th$ defined above, the remaining equation \eq{EqGang} is also satisfied.) Note that the general properties of~\eq{bsolalpha} are the same as described in the previous Sec.~\ref{Sec4.1} because this solution only differs with respect to~\eq{bsol2} by a multiplicative factor in front of the integral.

A drawback of this choice of equation of state is the occurrence of the term $1/A(r)$ in the expression of the tangential pressure~\eq{Pth1}, for $p_\th(r)$ is likely to diverge if the metric contains horizons. For example, with the simplest choice\footnote{This type of energy-momentum tensor is similar to the ones used in Refs.~\cite{Mureika:2010je,Gaete:2010sp} in the context of vector ungravity.} $\al = 0$, using the field equations one gets
\beq
\n{24}
\frac{A^\prime(r)}{A(r)} + B^\prime (r) = \frac{2 G  m(r)}{r^2 A(r)}.
\eeq
Thus, if the metric has a horizon at $r=r_H$, it follows $2Gm(r_H) = r_H > 0$, showing that the term in \eq{24} certainly diverges at the horizon. The only possibility for the tangential pressure [see~\eq{Pth1}]
\beq
\n{Pth_Alfa=0}
p_\th(r) = \frac{G m(r) \rho_{\rm eff}(r)}{2r A(r)}
\eeq
to be regular is if it happens that $\rho_{\rm eff}(r)$ vanishes at $r_H$. In general, the singularity of the tangential pressure at the horizons is related to the singularity of $B'(r)$ at $r=r_H$, see Eq.~\eq{bsolalpha}. These singularities are physical and cannot be removed by changing coordinates, for they also manifest in the curvature invariants.

One possible way to avoid the obstacles caused by the divergences at the horizons is to choose $\al = -1$. Indeed, for this particular value of the parameter, the effective tangential pressure [see~\eq{Pth1}],
\beq
\n{Pth_Alfa=-1}
p_\th  =  -    \rho_{\rm eff} - \frac{r \rho_{\rm eff}^\prime}{2}  ,
\eeq
is regular for any higher-derivative gravity model such that $f(-k^2)$ grows at least as fast as $k^4$ for sufficiently large $k$. 

However, the right-hand side of the equation~\eq{EqGrr} vanishes with this choice, so that $B = 0$ [see also Eq.~\eq{bsolalpha}]. Therefore, the metric \eq{metric} assumes the Schwarzschild-like form  
\beq
\label{metricB=0}
\rd s^2 = - A(r) \rd t^2 + \frac{\rd r^2}{A(r)} + r^2 \rd \Omega^2
.
\eeq
As mentioned in the Introduction, the disadvantage of \eq{metricB=0} is that, since $B = 0$, these types of solutions cannot reproduce the asymptotic Newtonian-limit behavior of the higher-derivative gravities.

Another starting point for obtaining this solution can be imposing \eq{metricB=0} as \textit{Ansatz} for the metric (which makes the system \eq{EqGtt}-\eq{EqGang} consistent) and then finding 
\beq
p_r = -\rho_{\rm eff}
\eeq 
as the solution of \eq{EqGrr}, and~\eq{Pth_Alfa=-1} as solution of~\eq{EqGang}. This approach was adopted in the frameworks of noncommutative gravity, generalized uncertainty principle, and other UV completions of gravity, e.g., in Refs.~\cite{Modesto:2010uh,Nicolini:2012eu,Isi:2013cxa,Nicolini:2019irw}, as well as in some local and nonlocal higher-derivative models~\cite{Modesto12,Bambi:2016wmo,Zhang14}.


\subsection{General case of regular pressures}
\label{SecPresGen}

As it was evident in the preceding discussion, the main difficulty in defining $B(r)$ is the presence of $A(r)$ in Eq.~\eq{EqGrr} if the metric has horizons. We also saw that the choice of radial pressure $p_r = - \rho_{\rm eff}$ was capable of curing pressure divergences at the horizons, with the side effect that $B \equiv 0$. Moreover, it was argued that the curvature singularity at the horizons was related to a divergence in the effective pressures. So, another possibility of regularizing these horizon physical singularities, but with a nontrivial $B(r)$, is with an equation of state in the form
\beq
\n{EOSSig}
p_r(r) = A(r) \si(r) - \rho_{\rm eff}(r) ,
\eeq
where $\si(x)$ is an $N_b$-regular continuous function (for an integer $N_b \geqslant 0$) such that the function $\xi(x) = x \si(x)$ is integrable on $[0,+\infty)$.

Indeed, this choice eliminates the term $1/A(r)$ in Eq.~\eq{Bsol}, so that the solution for $B(r)$ becomes 
\beq
\n{BsolSigma}
B(r) =  8\pi G \int_{\infty}^r \rd x \, x \, \si(x) 
.
\eeq
By definition, this integral is well defined for all values of $r$, and for convenience, we set $r_0 = \infty$. With the assumptions about $\si(x)$, it follows that $B(r)$ is $(N_b+1)$-regular.

To verify that the equation of state~\eq{EOSSig} makes the effective pressures regular, first we notice that if $\rho_{\rm eff}(r)$ is $N_a$-regular and $\si(x)$ is $N_b$-regular, then $A(r)$ is $(N_a+1)$-regular (see Sec.~\ref{Sec3}) and $p_r(r)$ is at least $N$-regular, with $N\equiv \min\lbrace N_a , N_b \rbrace$. In what concerns the tangential pressure defined by the conservation equation~\eq{Conserva}, 
\beq
\n{pthSig}
p_\th =  - \rho_{\rm eff} + A \si +  \frac{r}{4} \left(  3 A^\prime \si +  A B^\prime \si + 2 A \si^\prime -2 \rho_{\rm eff}^\prime \right),
\eeq
it is straightforward to verify that the order of regularity of each of the terms in the right-hand side of~\eq{pthSig} is either $N_a$ or $\min\lbrace N_a+1, N_b\rbrace$; therefore, $p_\th(r)$ is at least $N$-regular, just like $p_r(r)$.

Of course, in principle, there is a huge freedom to choose regular functions $\si(r)$ capable of yielding regular black hole solutions. For example, Ref.~\cite{dirty} in the context of noncommutative geometry used $\si(r) \propto r \rho_{\rm eff}(r)$. However, since for effective sources coming from 
higher-derivative gravity this function is only 0-regular, $B(r)$ cannot be more than 1-regular---regardless of the order of regularity of $\rho_{\rm eff}$. Further discussion about this choice of $\si(r)$ is carried out in Appendix~\ref{App1}. On the other hand, the multiplication for higher powers of $r$ can enhance the regularity of $B(r)$; for instance, if $\rho_{\rm eff}(r)$ is $N_a$-regular, the choice $\si(r) \propto r^{2n} \rho_{\rm eff}(r)$ (for a positive integer $n$) makes $B(r)$ to be $(N_a+n+1)$-regular. We shall consider another form for the function $\si(r)$ in what follows.


\subsection{Case $p_r = \left[  A(r) - 1 \right] \rho_{\rm eff} $}
\label{Sec-A}
\label{Sec4.4}

As a particular case of the equation of state presented in the previous subsection, let us consider a function $\si(r)$ such that the resultant pressure components have the following physical properties:
\begin{itemize}
\item[(1b)] The effective pressures vanish asymptotically for large $r$.
\item[(2b)] The effective pressures are finite at the horizons.
\item[(3b)] The effective pressures vanish at $r=0$.
\end{itemize}
While (1b) and (2b) are natural in the framework of Sec.~\ref{SecPresGen}, (3b) actually imposes restrictions on the form of $\si(r)$.\footnote{A slightly less stringent set of conditions was considered in~\cite{dirty}, with the specification for the effective pressures to be only finite (instead of vanishing) at $r=0$, like in Sec.~\ref{SecPresGen}.} The physical motivation for these requirements is to guarantee that the solution is asymptotically flat, everywhere regular and that it matches the small-$r$ behavior of the solution obtained in Sec.~\ref{Sec4.1} (which approximates the solution whose original source is a pointlike mass).

In the most interesting case (for this work) of a regular $\rho_{\rm eff}$, a simple equation of state satisfying the conditions (1b)--(3b) is
\beq
\n{EOS}
p_r(r) = \left[  A(r) - 1 \right] \rho_{\rm eff}(r),
\eeq
which follows from the choice $\si(r) = \rho_{\rm eff}(r)$ in the context of Sec.~\ref{SecPresGen}. Accordingly,
the solution~\eq{BsolSigma} for $B(r)$ reads 
\beq
\n{SolB-A}
B(r)  =  8 \pi G \int_\infty^r \rd x \, x \, \rho_{\rm eff} (x) ,
\eeq
while the tangential pressure~\eq{pthSig} is
\beq
\n{effpressth}
p_\th =  \left[  \frac{3}{4} A' r + A \left( 1 + 2 G \pi  r^2 \rho_{\rm eff} \right) - 1  \right] \rho_{\rm eff} \, + \, \frac{A - 1}{2} r \rho_{\rm eff}^\prime .
\eeq
It is straightforward to verify that~\eq{EOS} and~\eq{effpressth} indeed vanish as $r\to 0$ if $\rho_{\rm eff}$ is regular.\footnote{On the other hand, if $\rho_{\rm eff} \sim 1/r$, like in the case of fourth-derivative gravity [see Eq.~\eq{limitefonte4der}], the effective pressures tend to a nonzero value.} Also, 
from the general results of Sec.~\ref{SecPresGen}, we know that if $\rho_{\rm eff}$ is $N$-regular, $p_r$ and $p_\th$ are bounded and (at least) $N$-regular. A finer result is obtained in the case $\rho_{\rm eff}(r)$ is analytic\footnote{This is the case, for example, of the effective delta sources originated from local polynomial-derivative gravity and certain classes of nonlocal gravity~\cite{BreTib2,Nos6der}.} at $r=0$---in that case  $p_r$ and $p_\th$ are actually $(N+1)$-regular and, as a consequence, near the origin they are $O(r^2)$ even for $N=0$. This statement can be proved by direct calculation, using the Taylor series expansion of Eqs.~\eq{EOS} and~\eq{effpressth} [with~\eq{asol} and~\eq{massf}] and noticing the cancellation of the term of order $r^{2N+1}$.

From the features of $\rho_{\rm eff}$ mentioned in Sec.~\ref{Sec3}, it follows that 
the function $B(r)$ in~\eq{SolB-A} is bounded for any higher-derivative model. Moreover, 
from the considerations of Sec.~\ref{SecPresGen}, if $\rho_{\rm eff}$ is $N$-regular (or, equivalently, if $f(-k^2)$ grows at least as fast as $k^{4+2N}$ for a certain $N \geqslant 0$ and sufficiently large $k$), $B(r)$ is $(N+1)$-regular. In particular, the identity \eq{Bde} also holds, showing that the behavior of $B(r)$ in~\eq{SolB-A} near $r=0$ is very similar to the one of~\eq{bsol2}, as expected. 

Other choices of equations of state can also reproduce the same aforementioned physical effect on the solutions. Indeed,
this procedure has considerable ambiguity; for example, any equation of state in the form 
\beq
\n{EOS2}
p_r(r) = \left[  A^\ell(r) - 1 \right] \rho_{\rm eff}(r), 
\qquad \ell =1,2,\ldots ,
\eeq
following from $\si(r) = A^{\ell-1}(r) \rho_{\rm eff}(r)$, 
would also lead to the above properties (1b)--(3b). In this work, for the sake of simplicity, we adopt~\eq{EOS}.

\begin{center}
\vspace{+3pt}
$\ast$~~$\ast$~~$\ast$
\vspace{+3pt}
\end{center}

As a conclusion of this section, here we presented a recipe, based on Eq.~\eq{nlEE2}, of how to construct black hole solutions that are regular everywhere. First, choose a form factor that behaves as $f(-k^2) \sim k^4$ (or faster) for large $k$; then, choose effective pressures satisfying the conditions (1b) and (2b)---for example, through the function $\si(r)$ defined in Sec.~\ref{SecPresGen}. 

In addition, if $p_r$ and $p_\th$ vanish sufficiently fast for large $r$ and at the origin [condition (3b)], those solutions
reproduce the asymptotic behavior of the corresponding linearized higher-derivative model in the limits $r \to \infty$ and $r \to 0$. At the same time, near the horizons the metric is regularized by the regular pressures. One might even argue that these nontrivial pressure components can mimic the effect of the omitted higher-order terms when passing from Eq.~\eq{cu} to~\eq{nlEE2}. In our opinion, however, a proper address of this issue ultimately involves the study of the complete equations of motion, which exceeds the scope of this work.

Finally, for the equation of state~\eq{EOS} the metric is given by   
\beq
\label{metricB=0-UaiSo}
\rd s^2 = - \left(1-\frac{2G m(r)}{r} \right) \exp \left( 8 \pi G \int_\infty^r \rd x \, x \, \rho_{\rm eff} (x)  \right) \rd t^2 + {\left(1-\frac{2G m(r)}{r} \right)}^{-1} \rd r^2 + r^2 \rd \Omega^2
,
\eeq
with $m(r)$ defined in~\eq{massf}, as usual.


\section{Selected examples}
\label{Sec7}

In the previous sections we presented general results, which depended only on the asymptotic behavior of the form factor $F(z)$ of the model~\eqref{action}. This section aims to provide concrete examples where the results above can be explicitly verified and other particular features of the solutions discussed.   
In Sec.~\ref{Sec7.1} we present an example of nonlocal ghost-free gravity, while in Secs.~\ref{Sec7.15} and~\ref{Sec7.2} we discuss the cases of local higher-derivative gravity with simple poles. The general case of local higher-derivative gravity models, including degenerate poles, is considered in Sec.~\ref{Sec7.3}. A detailed analysis of the sixth-derivative Lee--Wick gravity is carried out in a separate publication~\cite{NosLW}.


\subsection{Nonlocal ghost-free gravity}
\label{Sec7.1}

\begin{figure}[b]
\includegraphics[scale=0.8]{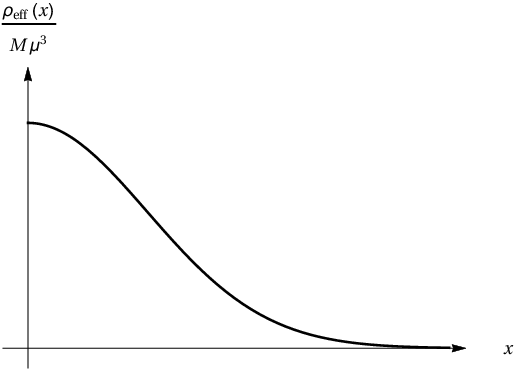}
\hspace{1.5cm}
\includegraphics[scale=0.8]{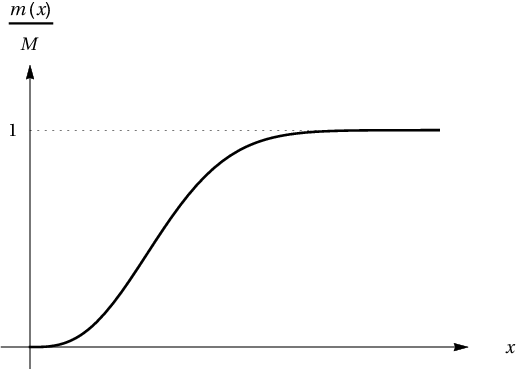}
\caption{Plot of the dimensionless forms of the effective delta source \eq{gauss} (left panel) and the mass function \eq{non-l-mass} (right panel) in terms of the variable $x = \mu r$. It is immediate to verify that the graphs are in agreement with Eqs.~\eq{rho-inf} and~\eq{rho-r0} for the effective source, and \eq{limM1} and \eq{limM2} for the mass function. Also, because $\rho_{\rm eff} (r)$ takes only positive values, the mass function is monotonic and $m(r) \leqslant M $; see the discussion around~\eq{implication}.}
\label{fig1}
\end{figure}

One of the simplest nonlocal higher-derivative models is defined by the form factor~\cite{Tseytlin:1995uq,SiegelEtAl1,SiegelEtAl2,SiegelEtAl3,Modesto12}
\beq
\n{ff-nonlocal-0}
F(z) = \frac{e^{- {z}/{\mu^2}} - 1}{z},
\eeq
for which
\beq
\n{ff-nonlocal-1}
f(z) = e^{- {z}/{\mu^2}}
,
\eeq
where $\mu>0$ has dimension of mass and defines the nonlocality scale. In this case, performing the integration~\eq{effsource}, the smeared effective delta source has a Gaussian profile~\cite{Tseytlin:1995uq},
\beq
\n{gauss}
\rho_{\rm eff} (r) = \frac{M \mu^3  }{8 \pi ^{3/2}} \, e^{-\frac{1}{4} \mu ^2 r^2},
\eeq
and the mass function is given by
\beq
\n{non-l-mass}
m(r) = M \left[ \text{erf}\left(\frac{\mu  r}{2}\right) - \frac{\mu r }{\sqrt{\pi }} \,  e^{-\frac{1}{4} \mu ^2 r^2}
\right]
.
\eeq
In Fig.~\ref{fig1} we show the behavior of the effective source and mass function, where the properties described in Sec.~\ref{Sec3} can be explicitly verified.

\begin{figure}[h]			
    \centering
    \begin{minipage}{0.45\textwidth}
        \centering
        \includegraphics[scale=0.7]{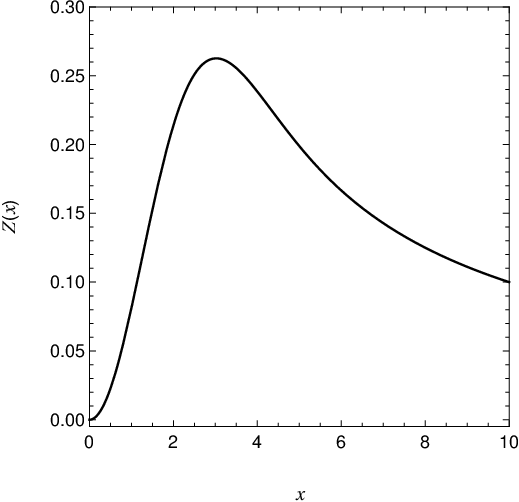}
\caption{Plot of $Z(x)$ in Eq.~\eq{q}.}
\label{fig2}
    \end{minipage}\hfill
    \begin{minipage}{0.45\textwidth}
        \centering
        \includegraphics[scale=0.7]{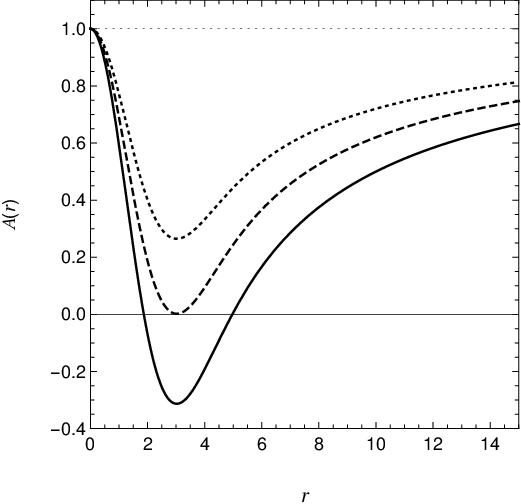}
\caption{Plot of $A(r)$ for $\mu = M_{\rm P} = 1$ with $M= 2.5 > M_{\rm c}$ (solid line, two horizons), $M= 1.9 = M_{\rm c}$ (dashed line, one extremal horizon) and $M = 1.4 < M_{\rm c}$ (dotted line, no horizon).}
\label{fig3}
    \end{minipage}
\end{figure}

Given \eq{non-l-mass}, the function $A(r)$ in~\eq{asol} reads
\beq
\label{Azao}
A(r) = 1 -\frac{2 G M }{r} \, \text{erf}\left(\frac{\mu  r}{2}\right)
+\frac{2 G M \mu }{\sqrt{\pi }} \, e^{-\frac{1}{4} \mu ^2 r^2}
.
\eeq
In order to evaluate the black hole mass gap, it is convenient to rewrite Eq.~\eq{Azao} in terms of the dimensionless function $Z(x)$, namely,
\beq
A(r) =  1 - 2GM \mu \, Z(\mu r),
\eeq 
where
\beq
\label{q}
Z(x) = \frac{1}{x}\, \text{erf}\left(\frac{x}{2}\right)-\frac{ e^{-\frac{x^2}{4}} }{\sqrt{\pi }}
.
\eeq
The plot of \eq{q} is shown in Fig.~\ref{fig2}, where one can verify that $Z(x)$ is bounded,
as expected from the general discussion in Sec.~\ref{Sec3}. Indeed, $\lim_{x \to 0} Z(x) = \lim_{x \to \infty} Z(x) = 0$, and it has an absolute maximum $Z_{\rm max} = 0.263$ at the point $x_* = 3.02$. Hence, the critical mass~\eq{Mcritial-gen} in this case is given by
\beq
M_{\rm c} = \frac{1}{2 G \mu\, Z_{\rm max}} = 1.9 \, \frac{M_{\rm P}^2}{\mu}
,
\eeq
where we used $G = 1/M_{\rm P}^2$.

Accordingly, if $M < M_{\rm c}$, $A(r)$ is always positive and the metric has no horizon. On the other hand, for $M > M_{\rm c}$, there exists a region where $A(r)$ takes negative values. Since the function $Z(x)$ in~\eq{q}  only has one local maximum, the equation $A(r) = 0$ can have at most two roots, $r_-$ and $r_+$, with $r_+ \geqslant r_-$. The values $r_{\pm}$ are the positions of the event horizon and an inner horizon. In Fig.~\ref{fig3} we show the plot of $A(r)$ for the two possible scenarios. Moreover, because $m(r) \leqslant M$, the position of the event horizon is bounded by the Schwarzschild radius, i.e.,
$
r_+ \leqslant r_{\rm s} = 2GM.
$

\subsubsection{Solution for $B(r)$ with $p_r = p_\th = 0$}
\label{subsubsub}

Firstly, let us analyze the framework of Sec.~\ref{Sec4.1}, i.e., with the effective pressures $p_r = p_\th = 0$,  corresponding to the case in which~\eq{nlEE2} is the small-curvature approximation for the full equations of motion of the theory~\eq{action}, sourced by a pointlike mass. The solution for $B(r)$ is obtained by substituting~\eq{gauss} and~\eq{Azao} in the integral~\eq{bsol2}, which can be evaluated numerically (if the integration interval does not cross a horizon). However, in order to study the curvature tensor components~\eq{CiDef} it suffices to consider the quantity
\beq
\label{Czao}
B'(r) = 8 \pi G \frac{ r \rho_{\rm eff} (r)}{A(r)},
\eeq
because the curvature components~\eq{CiDef} depend only on the derivatives of $B(r)$.

By using~\eq{gauss} and~\eq{Azao} into~\eq{Czao} we can evaluate each component~in~\eq{CiDef}. For simplicity, in what follows we focus only on the scalar curvature,
\beq
\n{RKKK}
R = 2 K_1 + 4 K_2 + 4 K_3 + 2 K_4, 
\eeq
but similar results also hold for the other invariants, such as the Kretschmann scalar~\eq{Kre}. The explicit expression for~\eq{RKKK} is
\beq
\label{RE}
R (r) = \frac{S(r)}{A(r)},
\eeq
where
\beq
\label{OEsse}
S(r) =  \, 
\frac{G  M  \mu^3  }{\sqrt{\pi }} \left[ 
1 
-\frac{3 G   M   }{ r} \, \text{erf}\left(\frac{\mu  r}{2}\right) \right]  e^{-\frac{1}{4} \mu ^2 r^2}
+\frac{3 G^2 M^2 \mu^4 }{\pi } \, e^{-\frac{1}{2} \mu ^2 r^2}.
\eeq
It is easy to verify that $\lim_{r \to \infty} R(r) = 0$ and 
\beq
\label{Rem0}
\lim_{r \to 0} R(r) = \frac{G M \mu^3 }{\sqrt{\pi }} .
\eeq
At intermediate scales, although $S(r)$ is a continuous, analytic even function of $r$, owning to the term $1/A(r)$ in~\eq{RE}, the behavior of $R(r)$ depends on the value of $M$ compared to the critical mass $M_{\rm c}$.

\begin{figure}[t]
\centering
\begin{subfigure}{.3\textwidth}
\centering
\includegraphics[scale=0.55]{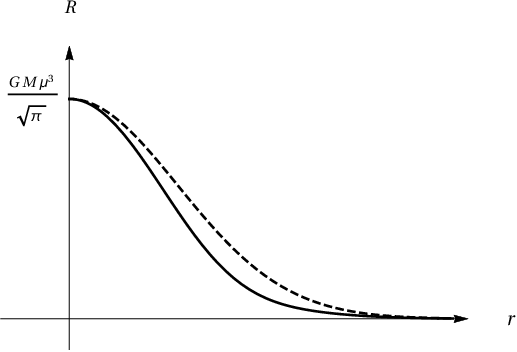}
\caption{\small Case $M < 0.66 M_{\rm c}$.}
\label{Fig4a}
\end{subfigure}%
\begin{subfigure}{.3\textwidth}
\centering
\includegraphics[scale=0.55]{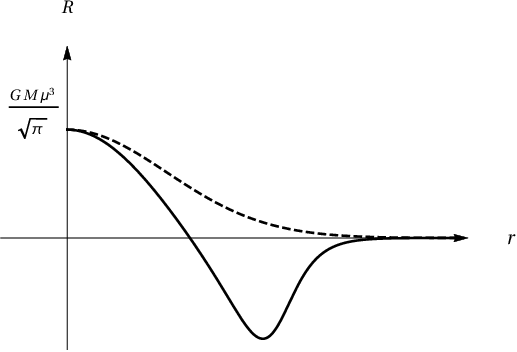}
\caption{\small Case $M \approx 0.95 M_{\rm c}$.}
\label{Fig4b}
\end{subfigure}
\begin{subfigure}{.3\textwidth}
\centering
\includegraphics[scale=0.55]{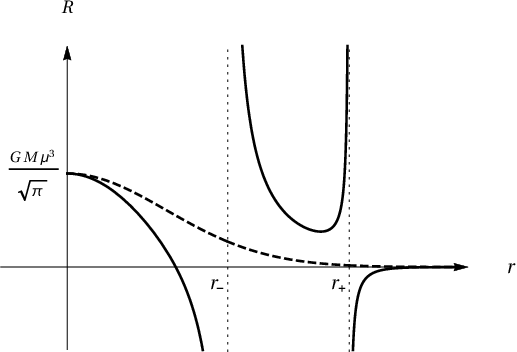}
\caption{\small Case $M > M_{\rm c}$.}
\label{Fig4c}
\end{subfigure}
\caption{Comparison between $R$ (solid line) and $R_{\rm \, lin}$ (dashed line), respectively given by Eqs.~\eq{RE} and~\eq{R-lin}. The panel~(a) shows the case in which the mass is considerably smaller than the critical mass: $R$ is positive and achieves its maximum at $r=0$; the two solutions have the same qualitative behavior. For larger values of the mass (but still smaller than $M_{\rm c}$), $R$ may achieve arbitrarily large negative values [panel (b)]. The maximum of $\vert R \vert$ can occur at $r \neq 0$, which shows a significant difference between the two solutions; this marks the approximation breakdown. The panel (c) illustrates the typical case when $M>M_{\rm c}$, for which $R$ diverges at the horizons.}
\label{Fig4}
\end{figure}

In Fig.~\ref{Fig4} we show the typical behavior of~\eq{RE} for $M < M_{\rm c}$ and $M > M_{\rm c}$. We also compare both graphs with the Newtonian-limit solution~\eq{metric-New}, with the potential associated to the form factor~\eq{ff-nonlocal-1}~\cite{Tseytlin:1995uq,SiegelEtAl1,SiegelEtAl2,SiegelEtAl3,Modesto12},
\beq
\n{maz_sqn}
\ph (r) = - \frac{GM}{r} \, {\rm erf} \left( \frac{\mu r}{2} \right),
\eeq
for which the linearized Ricci scalar reads 
\beq
\n{R-lin}
R_{\rm \, lin} = \frac{G \mu ^3 M }{\sqrt{\pi }} \, e^{-\frac{1}{4} \mu ^2 r^2}
.
\eeq  
As one can see from the graphs, in both cases, the linearized and nonlinearized solutions match pretty well for very small and large $r$, i.e., for $r \ll 1/\mu$ and $r \gg 1/\mu$.
Also, if $ M$ is sufficiently small, the two solutions do not deviate too much; we elaborate more on this in what follows.

As discussed in Sec.~\ref{Sec4}, if $M < M_{\rm c}$ we have $A(r) > 0 $ everywhere and the scalar curvature \eq{RE} is bounded. To be more precise, if $M < 0.66 M_{\rm c}$ the function $S(r)$ is positive and $R$ achieves the maximum value at $r=0$, see Eq.~\eq{Rem0} and Fig.~\ref{Fig4a}. For larger values of $M$ within the range $0.66 M_{\rm c} < M < M_{\rm c}$, the negative term in~\eq{OEsse} becomes more relevant. It dominates for $0.95 M_{\rm c} < M < M_{\rm c}$ (Fig.~\ref{Fig4b})---in this range, $\vert R(r) \vert$ can assume arbitrarily large values as $M$ approaches the critical mass. 
In fact, for $M > M_{\rm c}$ the equation $A(r) = 0$ has two roots, and the scalar curvature diverges at the horizon positions $r_{\pm}$ (see Fig.~\ref{Fig4c}). Therefore, combining these results with the numeric factors in Eq.~\eq{Rem0} one can show that, even though $R$ is bounded for any value of $M$ smaller than $M_{\rm c}$, the small-curvature condition stated in the form $\vert R \vert/\mu^2 < 1$ actually requires a slightly more stringent constraint, namely, $M < 0.93  M_{\rm c}$.

Analogous results can be formulated for the other curvature-squared invariants, $R_{\mu\nu}^2$ and $R_{\mu\nu\al\beta}^2$. This explicitly shows that there exists a value for the mass $M<M_{\rm c}$ such that the small-curvature condition $\mathcal{R}^2 /\mu^4 < 1$ holds ($\mathcal{R}$ is a generic curvature tensor). The $O(R^2)$ corrections in~\eq{cu} might be irrelevant in these circumstances. We emphasize, however, that the approximation holds \emph{for sufficiently small} $r$ even when $M > M_{\rm c}$ because the nonlinearities in curvatures are suppressed in this limit, as discussed in general grounds in Sec.~\ref{singu}.

As explained before, since the form factor~\eq{ff-nonlocal-1} grows faster than any polynomial, the metric components $A(r)$ and $B(r)$ are even functions. This makes all curvature invariants to be singularity free at the origin, even those constructed with derivatives of the curvature tensors. Below, we present some values of these invariants at $r=0$ [see also~\eq{Rem0}]:
\beq
\label{KeRic2}
R_{\mu\nu\al\beta} R^{\mu\nu\al\beta} (0) = \frac{5 G^2  M^2 \mu ^6}{3 \pi }
,
\qquad
R_{\mu\nu} R^{\mu\nu} (0) = \frac{G^2  M^2 \mu ^6}{\pi }
,
\eeq
\beq
\label{RiemBoxRiem1}
\Box R (0) = -\frac{G  M \mu ^5}{2 \pi } \left(3 \sqrt{\pi } + 2 G \mu  M\right)
,
\qquad
\Box^2 R (0) =  \frac{ G  M \mu ^7 }{12 \pi ^{3/2}} \big( 45 \pi  +138 \sqrt{\pi } G \mu  M -52 G^2 \mu ^2 M^2 \big)
.
\eeq

It is remarkable that with the very simple equation~\eq{nlEE2}, one can obtain the same qualitative results of~\cite{Holdom:2002xy}.
Namely, the curvatures are regular at $r = 0$ and weak around the origin, there exists a mass gap for the black hole formation, and the curvatures abruptly grow close to the horizon positions (if there are horizons).


\subsubsection{Solution for $B(r)$ with $p_r = [A(r) - 1] \rho_{\rm eff}$}

Let us now obtain the solution associated with the equation of state~\eq{EOS}, of Sec.~\ref{Sec4.4}, which regularizes the divergences of the pressures at the horizons. The quadrature~\eq{SolB-A} for $B(r)$ can be explicitly evaluated in this case,
\beq
B(r) = -\frac{2 G \mu  M }{\sqrt{\pi }} \, e^{-\frac{1}{4} \mu ^2 r^2}
.
\eeq 
So, the metric is given by
\beq
\begin{split}
\label{metricB=0-exp}
\rd s^2 = &- \left[1 -\frac{2 G M }{r} \, \text{erf}\left(\frac{\mu  r}{2}\right)
+\frac{2 G M \mu }{\sqrt{\pi }} \, e^{-\frac{1}{4} \mu ^2 r^2} \right] \exp \left( -\frac{2 G \mu  M }{\sqrt{\pi }} \, e^{-\frac{1}{4} \mu ^2 r^2}  \right) \rd t^2 
\\
&
+ \left[ 1 -\frac{2 G M }{r} \, \text{erf}\left(\frac{\mu  r}{2}\right)
+\frac{2 G M \mu }{\sqrt{\pi }} \, e^{-\frac{1}{4} \mu ^2 r^2} \right]^{-1} \rd r^2 + r^2 \rd \Omega^2
,
\end{split}
\eeq
which is consistent with the Newtonian approximation described in Sec.~\ref{Sec5}. Indeed, expanding the temporal part of~\eq{metricB=0-exp} in powers of $M$, at leading order we get $g_{tt} = - [1+2\ph(r)]$, with $\ph(r)$ given by~\eq{maz_sqn}.

\begin{figure}[h]
\includegraphics[scale=0.85]{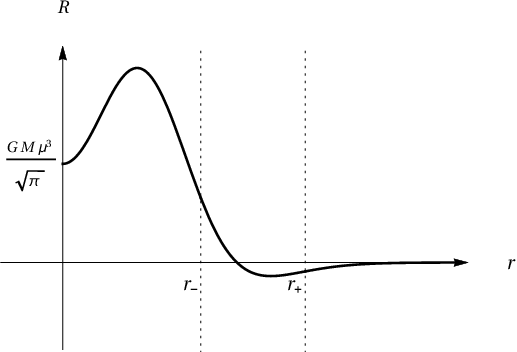}
\caption{\small Behavior of the Ricci scalar for the metric~\eq{metricB=0-exp} when $M > M_c$. The scalar curvature is finite everywhere, even at the horizons $r_\pm$.}
\label{Fig5}
\end{figure}

In consonance with the discussion of Sec.~\ref{Sec6}, in this case the curvature invariants are regular everywhere. Indeed, in Fig.~\ref{Fig5} we show the typical behavior of the scalar curvature $R\,$ for $M > M_{\rm c}$; comparing with Fig.~\ref{Fig4c}, we notice that this choice of effective pressures regularizes the singularities at the horizons. Moreover, the values of the curvature invariants at $r=0$ are the same as in Eqs.~\eq{Rem0} and \eq{KeRic2},
because both equations of state lead to solutions that have the same behavior for sufficiently small~$r$, as explained in Sec.~\ref{Sec4.4}. (However, curvature-derivative invariants calculated at the two solutions might differ, as they involve higher-order metric derivatives.) 
 We do not plot the case where $M < M_{\rm c}$ since it is very similar to the one shown in Fig.~\ref{Fig4a}.

Finally, we point out that the metric~\eq{metricB=0-exp} differs from the one obtained in Ref.~\cite{dirty} inspired by noncommutative geometry. Although both come from the same Gaussian effective source~\eq{gauss}, the underlying equations of state are different, which leads to different shift functions $B(r)$. In Appendix~\ref{App1}, we carry out a more detailed comparison of the two solutions.

\subsection{Sixth-derivative gravity with real poles}
\label{Sec7.15}

As a second example, let us consider the case where the form factor $f(z)$ is a polynomial of degree two with simple real roots $m_1^2$ and $m_2^2$, namely,
\beq
\label{fsixd}
f(z) = \frac{(m_1^2 - z) (m_2^2 - z) }{m_1^2 m_2^2} ,
\eeq
which corresponds to a sixth-derivative gravitational action. The effective delta source~\eq{effsource} for this model was calculated in~\cite{BreTib2} and it reads
\beq
\label{rhosixd}
\rho_{\rm eff} (r) = \frac{M m_1^2 m_2^2}{4 \pi \left( m_2^2 - m_1^2 \right) }  \frac{e^{-m_1 r} - e^{-m_2 r}}{r}   .
\eeq
Ordering the masses such that $m_1 < m_2$, it is easy to see that $\rho_{\rm eff} (r) > 0$ for all $r$. According to the discussion involving Eq.~\eq{implication}, it follows that the mass function, 
\beq
\label{msixd}
m(r) = M \left[ 1 + \frac{m_1^2(1+m_2 r)e^{-m_2 r} - m_2^2 (1+m_1 r)e^{-m_1 r}}{m_2^2 - m_1^2}   \right] 
,
\eeq
is monotone and also positive; this can be explicitly verified by checking that $m^\prime(r)>0$ and $m(0)=0$. 

It is worth mentioning that the limit $m_2 \to m_1$ of the expressions~\eq{rhosixd} and~\eq{msixd} is smooth, and it results in the formulas for the case of sixth-derivative gravity with degenerate masses, namely,
\beq
\label{den}
\rho_{\rm eff} (r)\big\vert_{m_2=m_1} = \frac{M m_1^3}{8 \pi } e^{-m_1 r} ,  \qquad
m(r)\big\vert_{m_2=m_1} = M \left[ 1 - \left( 1 + m_1 r + \frac{1}{2} m_1^2 r^2 \right)  e^{-m_1 r}  \right] .
\eeq
Since the UV behavior of the form factor does not depend on the multiplicity of the roots, the effective sources in~\eq{den} and~\eq{rhosixd} have the same order of regularity, as it can be easily checked by comparing their Taylor series. For further discussion regarding the limit $m_2 \to m_1$ in linearized sixth-derivative gravity, see~\cite{Accioly:2016qeb}.

\begin{figure}[t]
\includegraphics[scale=0.7]{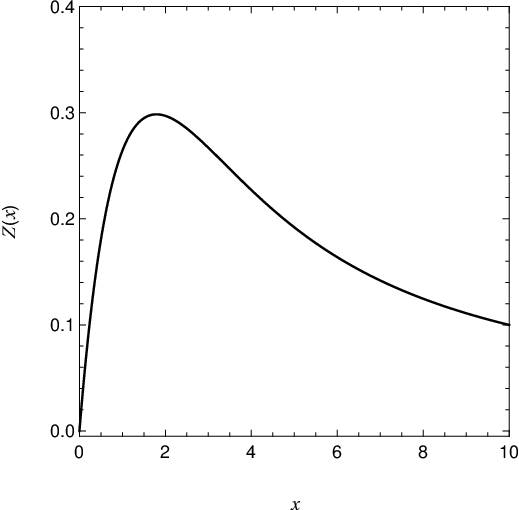}
\caption{Graph of $Z(x)$ in Eq.~\eqref{ZndPoly}. The function is positive for every $x>0$  and it has a maximum 
$Z_{\rm max} = 0.298$ at $x_* = 1.79$. \label{Fig6derZ}}
\end{figure} 

The expression for $A(r)$ is obtained from combining~\eq{asol} and \eqref{msixd}, and it can be cast in the form\footnote{In particular, in the degenerate case $m_2 = m_1$ we obtain
$$
A(r)\big\vert_{m_2=m_1} = 1 - \tfrac{2 G M }{r} \left[  1 - \left( 1 + m_1 r + \tfrac{1}{2} m_1^2 r^2 \right)  e^{-m_1 r} \right].
$$}
\beq \label{Andsixd}
A(r) = 1 - 2GM  \frac{ m_1 m_2}{m_2^2 - m_1^2} \left[ m_2 Z(m_1 r)  - m_1 Z(m_2 r) \right] ,
\eeq
where  we defined the dimensionless function 
\beq\label{ZndPoly}
Z(x)=\frac{1}{x}[1-(1+x)e^{-x}].
\eeq
Since $Z(x)$ is bounded (see Fig.~\ref{Fig6derZ}), $A(r)$ is also bounded. Moreover, 
the function $Z(x)$ is positive for any $x>0$  and takes the maximal value
$Z_{\rm max} = 0.298$ for $x_* = 1.79$. Therefore, 
\beq
\max_r \left\lbrace m_2 Z(m_1 r)  - m_1 Z(m_2 r) \right\rbrace \leqslant \max_r \left\lbrace m_2 \vert Z(m_1 r) \vert  + m_1 \vert Z(m_2 r)\vert \right\rbrace \leqslant (m_1 + m_2) Z_{\rm max} ,
\eeq
which leads to the conclusion that $A(r)$ is positive if $M$ is small enough such that
\beq
\label{MassGapSix}
2GM  \frac{ m_1 m_2}{m_2 - m_1} Z_{\rm max}  < 1 .
\eeq
This yields a lower-bound estimate to the critical mass $M_{\rm c}$ of Eq.~\eq{Mcritial-gen},
\beq
\label{LBEst}
M_{\rm c} \geqslant 1.68 \, \frac{M_{\rm P}^2}{  \tilde{\mu}}   , \qquad \tilde{\mu} \equiv \frac{ m_1 m_2}{m_2 - m_1}, 
\eeq
where again we used $G = 1/M_{\rm P}^2$. We remark that $M_{\rm c}$ might actually be considerably bigger than the estimate~\eq{LBEst}, which is only a lower bound. Nevertheless, if $M < 1.68 M_{\rm P}^2 / \tilde{\mu}$ the solution describes a horizonless compact object regardless of the choice of the equation of state in the framework of Sec.~\ref{SecPresGen}.

\begin{figure}[t]
\includegraphics[scale=0.7]{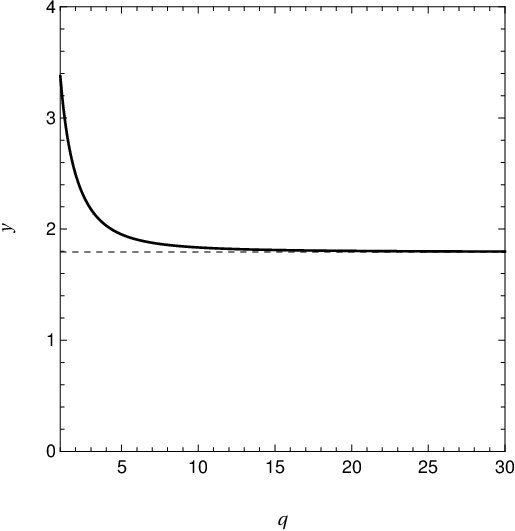}
\caption{Numerical solution of Eq.~\eqref{G=0}, which shows that 
it has only one solution 
for each $q \geqslant 1$. In the limit $q \to 1$ the equation is solved for $y = 3.38$, while for large $q$ the solution for $y$ tends to 1.79.
}
\label{Fig6der}
\end{figure}

On the other hand, for $M > M_{\rm c}$, the solution has exactly \emph{two} horizons. To prove this statement, it suffices to show that the function $A(r)$ in Eq.~\eq{Andsixd} has only one local minimum. In particular, this is true if the equation $A^\prime(r) = 0$ has only one solution\footnote{The point where $A^\prime(r) = 0$ is a minimum, and not a maximum, because~\eq{Andsixd} gives $A^{\prime\prime}(0)= - \tfrac{4}{3} G m_1^2 m_2^2 / (m_1+m_2)<0$, i.e., $A(r)$ is concave downward near $r=0$.} for a given pair of masses $m_1$ and $m_2$. If we define the dimensionless variables
\beq
q \equiv \frac{m_2}{m_1}  \qquad \text{and} \qquad y \equiv m_1 r,
\eeq
the equation $A^\prime(r) = 0$ can be cast as
\beq
\label{G=0}
\frac{1-q^2+q^2  \left(1+y+y^2\right)e^{-y} - \left[ 1+q y (1+q y)\right]e^{-q y}  }{1-q^2} 
= 0.
\eeq
The numerical solution of this equation in the $qy$-plane is shown in Fig.~\ref{Fig6der}, where one can see that it has only one solution for each value of $q$. In other words, given a pair $(m_1,q)$---or, equivalently, $(m_1,m_2)$---the function $A(r)$ has only one minimum.

Figure~\ref{Fig6der} can also be regarded as the graph of a function $y_\text{min}(q)$, which is related to the position $r_{*}$ of the minimum\footnote{The quantity $r_{*}$ defined here is the same one that appeared in the general discussion of Sec.~\ref{Sec3} [see Eq.~\eq{Mcritial-gen}], because the minimum of $A(r)$ corresponds to the maximum of $\tilde{m}(r)/r$.} 
of $A(r)$, namely, 
\beq
r_{*} = \frac{y_\text{min}(q)}{m_1} . 
\eeq
Therefore, in Fig.~\ref{Fig6der} we can identify two special distinct regimes, namely, $q \to 1$ and $q\to \infty$. The former corresponds to the degenerate case $m_1 = m_2$, for which $y_\text{min}$ assumes its maximal value $y_\text{min}(1)=3.38$. As $q$ increases, the function $y_\text{min}(q)$ monotonically decreases, and it tends to the constant value $ x_* = 1.79 $ [the same position of the extremum of $Z(x)$, see Eq.~\eq{ZndPoly}]. This latter regime corresponds to the case in which $m_2 \gg m_1$, i.e., the massive mode $m_2$ is too heavy and has a very short range; consequently, for larger distances, the model behaves similarly to the fourth-derivative gravity with only the scale $m_1$. Both situations are analogous to what was described in~\cite{Accioly:2016qeb} about the weak-field limit of sixth-derivative gravity. Since the outer (event) horizon $r_+$ is bounded from below by $r_*$, and by the Schwarzschild radius from above [see Eq.~\eq{bound}], we conclude that
\beq
\frac{1.79}{m_1} \leqslant r_+ \leqslant 2 G M .
\eeq

Let us stress that the case of sixth-derivative gravity in which the masses $m_1 \neq m_2$ in~\eq{fsixd} are complex conjugate is completely different from the one studied here, for the effective source is not positive and the mass function has oscillations. For instance, this allows the metric to have \emph{more} than two horizons, as we showed in detail in~\cite{NosLW}.

In what concerns the nontrivial possibilities for $B(r)$, let us only discuss the solution with regular pressures in the framework of Sec.~\ref{Sec4.4}, i.e., with the equation of state~\eq{EOS} (the case of null pressures is very similar to Sec.~\ref{subsubsub}). In this case, Eq.~\eq{SolB-A} yields\footnote{In the degenerate scenario $m_2 = m_1$ this expression reduces to
$$
B(r)\big\vert_{m_2=m_1}  = - G M m_1 (1 + m_1 r) e^{- m_1 r}.
$$}
\beq\label{B_real_6der}
B(r)  = -\frac{2 G M m_1 m_2 }{m_2^2 - m_1^2} \left( m_2 e^{-m_1 r} -  m_1 e^{-m_2 r} \right) .
\eeq
Combining \eqref{Andsixd} and \eqref{B_real_6der} we obtain the explicit expression for the metric,
\beq
\begin{split}
\label{metricB-6der}
\rd s^2 = &- \left\lbrace 1 - 2GM  \frac{ m_1 m_2}{m_2^2 - m_1^2} \left[ m_2 Z(m_1 r)  - m_1 Z(m_2 r) \right] \right\rbrace 
\exp \left[ -\frac{2 G M m_1 m_2 }{m_2^2 - m_1^2} \left( m_2 e^{-m_1 r} -  m_1 e^{-m_2 r} \right)  \right] \rd t^2 
\\
&
+ \left\lbrace 1 - 2GM  \frac{ m_1 m_2}{m_2^2 - m_1^2} \left[ m_2 Z(m_1 r)  - m_1 Z(m_2 r) \right] \right\rbrace ^{-1} \rd r^2 + r^2 \rd \Omega^2
.
\end{split}
\eeq
It is easy to check that to the leading order in $M$, the above metric reduces to the Newtonian-limit form~\eq{metric-New}, 
where 
\beq
\ph(r) = - \frac{G M}{r} \left( 1 - \frac{m_2^2 \, e^{-m_1 r} - m_1^2 \, e^{-m_2 r}}{m_2^2 - m_1^2} \right)
\eeq
is precisely the modified Newtonian potential~\cite{Newton-MNS,Accioly:2016qeb}, in agreement with the general discussion of Sec.~\ref{Sec5}.


\subsection{Polynomial gravity with simple real poles}
\label{Sec7.2}

As a more general but less explicit example, let us consider the case where the form factor $f(z)$ is a generic polynomial of degree $N \geqslant 2$ with real distinct roots, i.e.,
\beq
\n{fz-poly}
f(z) =1+zF(z)= \prod_{i=1}^N \frac{m_i^2 - z}{m_i^2} \; ,
\eeq
with $m_i\neq m_j$ if $i\neq j$. (The more general case in which the roots can be complex and/or degenerate is discussed in Sec.~\ref{Sec7.3} below.)
The expression for the effective source can be obtained by expanding the integrand of~\eq{effsource} in partial fractions, using 
\beq \label{PartFrac1}
\frac{1}{f(z)} =  \sum_{i=1}^N C_{i} \frac{ m_i^2}{m_i^2 - z}  ,
\eeq
where the coefficients $C_{i}$ are given by
\beq
\n{coi}
C_{i} =  \prod_{\substack{j=1\\j \neq i}}^N \frac{m_j^2}{m_j^2 - m_i^2}  
\eeq
and satisfy the identities~\cite{Nos6der,PU50,QuandtSchmidt}
\beq
\label{gidND}
\sum_{i=1}^N C_{i}\; m_i^{2k} =
\begin{cases}
1, \qquad 
\text{for $k=0$,}\\
0,
\qquad 
\text{for $k=1,2,\dots, N -1$}.
\end{cases}
\eeq

Substituting~\eq{PartFrac1} into~\eqref{effsource} it follows~\cite{BreTib2}
\beq
\n{rho_real_poly}
\rho_{\text{eff}}(r)=
\frac{M}{4 \pi r }\sum_{i=1}^{N}C_{i} \, m_{i}^2 e^{-m_i r}.
\eeq
Hence, the effective mass function~\eqref{massf} reads
\beq 
\n{mass_real_poly}
m(r)=M \left[ 1-\sum_{i=1}^N C_{i}\;(1+m_i r)e^{-m_i r}  \right] .
\eeq
We did not encounter any effective source~\eq{rho_real_poly} that assumes negative values; therefore, we conjecture that $\rho_{\text{eff}}(r)$ above is strictly positive, which makes $m(r)$ to be a monotonic function and the horizons to be bounded by the Schwarzschild radius, according to the discussion involving Eq.~\eq{implication}. 
Let us remark once more that the situation with complex quantities $m_i^2$ is entirely different, and in that case $\rho_{\text{eff}}(r)$ can achieve negative values, with important implications for the black hole solutions~\cite{NosLW}.

The combination of~\eq{asol} and \eqref{mass_real_poly} yields
\beq
\n{A_real_poly}
\begin{split}
A(r)= 1-\frac{2GM}{r}\; \left[  1-\sum_{i=1}^N C_{i}\;(1+m_i r)e^{-m_i r} \right]
.
\end{split}
\eeq
From the identity~\eqref{gidND} with $k=0$, it is easy to see that the property~\eq{ALim} is satisfied, in particular, 
$\lim_{r\to 0} A(r) = 1 $.
The equation~\eqref{A_real_poly} can be rewritten in terms of the same dimensionless function $Z(x)$ in~\eq{ZndPoly},
\beq \label{AndPloy}
A(r)=1 - 2GM\; \sum_{i=1}^N C_{i} \, m_i \, Z(m_i r) .
\eeq
Since $Z(x)$ is bounded by $0 \leqslant Z(x) \leqslant Z_{\rm max} = 0.298$, $A(r)$ is also bounded. Therefore, the same reasoning we adopted to estimate the mass gap in Sec.~\ref{Sec7.15} leads to the conclusion that 
$A(r)$ is positive provided that $M$ is small enough such that
\beq
\label{MassGap}
2GM  \tilde{\mu} Z_{\rm max}   < 1 ,
\eeq
where
\beq\label{curlyM}
\tilde{\mu} \equiv \sum_{i=1}^N \vert C_{i} \vert m_i  .
\eeq
This yields the lower-bound estimate to the critical mass $M_{\rm c}$ in Eq.~\eq{Mcritial-gen},
\beq
\label{LBEst2}
M_{\rm c} \geqslant 1.68 \, \frac{M_{\rm P}^2}{  \tilde{\mu}}   .
\eeq
Notice that Eqs.~\eq{LBEst2} and~\eq{LBEst} are essentially the same, the only difference being the mass scale $\tilde{\mu}$. This is expected, for the previous section's model is the particular case $N=2$ of the model considered here.

Finally, the equation of state~\eq{EOS} in the framework of Sec.~\ref{Sec4.4} results in the solution~\eq{SolB-A} for $B(r)$, namely, 
\beq\label{B_real_poly}
B(r)  = -2GM \sum_{i=1}^N C_{i} \, m_i \, e^{- m_i r}.
\eeq
The complete metric, thus, reads
\beq
\begin{split}
\label{metricB=0-polys}
\rd s^2 = &- \left\lbrace 1-\frac{2GM}{r}\; \Big[  1-\sum_{i=1}^N C_{i}\;(1+m_i r)e^{-m_i r} \Big] \right\rbrace 
\exp \left( -2GM \sum_{i=1}^N C_{i} \, m_i \, e^{- m_i r}  \right) \rd t^2 
\\
&
+ \left\lbrace 1-\frac{2GM}{r}\; \Big[  1-\sum_{i=1}^N C_{i}\;(1+m_i r)e^{-m_i r} \Big] \right\rbrace ^{-1} \rd r^2 + r^2 \rd \Omega^2
.
\end{split}
\eeq
In agreement with the general discussion of Sec.~\ref{Sec5}, in the leading order in $M$ this metric reduces to the Newtonian form~\eq{metric-New} with
\beq
\ph(r) = - \frac{G M}{r} \left( 1 + \sum_{i=1}^N C_{i} \, e^{-m_i r}  \right)
\eeq
being exactly the modified Newtonian potential evaluated in~\cite{Newton-MNS}.


\subsection{General polynomial gravity}
\label{Sec7.3}

Finally, let us consider the most general case of polynomial gravity, where we allow the mass parameters $m_i $ to be complex and/or degenerate. The form factor reads
\beq
f(z)=\prod_{i=1}^{n} \left( \frac{m_i^2 - z}{m_i^2}\right)^{\alpha_i} ,
\eeq
where we assume that the polynomial $f(z)$ has degree $N \geqslant 2$ and it has $n \geqslant 1$ distinct roots
$m_1^2, m_2^2, \dots, m_n^2$, each of which has a 
multiplicity $ \alpha_1, \alpha_2, \dots, \alpha_n$, such that 
\beq
N=\sum_{i=1}^{n}\alpha_i \; .
\eeq

Similarly to the previous example, we can use the partial fraction decomposition to write
\beq
\label{PFD1}
\frac{1}{f(z)}= \sum_{i=1}^{n} \sum_{j=1}^{\alpha_i}\mathcal{C}_{i,j} \left( \frac{ m_i^{2} }{m_i^2 -z }\right) ^j ,
\eeq
where the coefficients can be determined by the residue method (see, e.g.,~\cite{Grad}) and are given by
\beq\label{G_coeff_C}
\mathcal{C}_{i,j}= \frac{(-1)^{\al_i-j}}{m_i^{2j} \, (\alpha_i -j)!} \, \frac{\rd^{\alpha_i -j}}{\rd z^{\alpha_i -j} }
\left. \frac{(m_i^2 - z)^{\alpha_i}}{f(z)}\right|_{z=m_i^2} .
\eeq
These coefficients satisfy identities that generalize the ones in~\eqref{gidND}, in particular,
\beq
\label{Ids2}
\sum_{i=1}^n \sum_{j=1}^{\alpha_i}\mathcal{C}_{i,j} = 1  \qquad \text{and} \qquad \sum_{i=1}^n \mathcal{C}_{i,1} m_i^2 = 0 .
\eeq
Calculating the effective delta source $\rho_\text{eff}$ from~\eq{effsource} using~\eq{PFD1}, we find~\cite{BreTib2}
\beq
\begin{split}\label{rhoC}
\rho_\text{eff}(r)= \frac{M}{4 \pi^{3/2}}  \sum_{i=1}^{n} \sum_{j=1}^{\alpha_i} \, \mathcal{C}_{i,j} \frac{m_i^{2j}}{(j-1)!} 
\left(\frac{r}{2 m_i} \right)^{j-\frac{3}{2}} K_{j-\frac{3}{2}} (m_i r) ,
\end{split}
\eeq
where $K_\nu$ is the modified Bessel function of the second kind~\cite{Grad}. As shown in~\cite{BreTib2}, the latter identity in~\eq{Ids2} is responsible for the 0-regularity of the above effective source.

It is also useful to write explicitly the expression of the modified Newtonian potential $\ph(r)$, obtained from the modified Poisson equation~\eqref{new-pot-eq-1} and calculated in~\cite{BreTib1},
\beq \label{phiA}
\begin{split}
\varphi(r)=-\frac{G M}{r} \left[ 1 - \frac{2}{\sqrt{\pi}} \sum_{i=1}^{n} \sum_{j=1}^{\alpha_i}  \frac{\mathcal{A}_{i,j}}{(j-1)!} \left(\frac{m_i r}{2} \right)^{j-\frac{1}{2}} K_{j-\frac{3}{2}} (m_i r)\right] \, ,
\end{split}
\eeq
where the coefficients $\mathcal{A}_{i,j}$ are defined such that\footnote{Comparing to the notation used in~\cite{BreTib1}, the correspondence between the coefficients $A_{i,j}$ (used there) and $\mathcal{A}_{i,j}$ (used here) is through $\mathcal{A}_{i,j} = (j - 1)! \, A_{i,j}/m_i^{2(j-1)}$.}
\beq
\label{PFD2}
\frac{1}{z f(z)} = \frac{1}{z} + \sum_{i=1}^{n} \sum_{j=1}^{\alpha_i} \frac{\mathcal{A}_{i,j}}{m_i^2} \left(  \frac{ m_i^2 }{m_i^2 -z } \right) ^j ,
\eeq
namely,
\beq
\label{Aij}
\mathcal{A}_{i,j} = \frac{(-1)^{\al_i-j}}{m_i^{2(j-1)} \, (\alpha_i - j)!} \, \frac{\rd^{\alpha_i -j}}{\rd z^{\alpha_i-j}} \frac{\left( m_i^2 - z\right)^{\alpha_i}}{z f(z)}  \Bigg|_{z = m_i^2} \, .
\eeq
Hence, comparing~\eq{PFD1} and~\eq{PFD2} one can show that
\beq \label{relationAC}
\mathcal{A}_{i,j} =  \sum_{k=j}^{\alpha_i} \mathcal{C}_{i,k}   .
\eeq

At this point, we can use relations~\eq{alinsol},~\eq{aebf} and~\eq{phiA} to obtain compact expressions for the mass function and the function~$A(r)$, 
\beq\label{mass_C}
\begin{split}
m(r)=M \left[ 1-\frac{4 }{\sqrt{\pi}} \sum_{i=1}^{n} \sum_{j=1}^{\alpha_i}  \frac{ \mathcal{A}_{i,j} }{(j-1)!}   
 \left(\frac{m_i r }{2} \right)^{j+\frac{1}{2}} K_{j-\frac{5}{2}} (m_i r)     \right]  \, ,
\end{split}
\eeq
\beq\label{A_C}
\begin{split}
A(r)=1-\frac{2GM}{r} +\frac{8 G M }{\sqrt{\pi} r } \sum_{i=1}^{n} \sum_{j=1}^{\alpha_i} 
\frac{\mathcal{A}_{i,j}}{(j-1)!} 
 \left(\frac{m_i r }{2} \right)^{j+\frac{1}{2}} K_{j-\frac{5}{2}} (m_i r) \, .
\end{split}
\eeq
One can easily check that the formulas \eqref{phiA}, \eqref{mass_C}, and \eqref{A_C} reduce to the simple pole ones of Sec.~\ref{Sec7.2} if we set $\alpha_i=1 \,\forall\, i$. For instance, in this case, \eqref{G_coeff_C} coincides with~\eqref{coi}.

Finally let us consider the solution described in Sec.~\ref{Sec4.4}, with an effective pressure defined by the equation of state \eqref{EOS}. From Eq.~\eqref{SolB-A}, we obtain the nontrivial shift function
\beq\label{B_A}
\begin{split}
B(r)= - \frac{4 G M}{ \sqrt{\pi}}  \sum_{i=1}^{n} \sum_{j=1}^{\alpha_i}  \frac{   \mathcal{C}_{i,j} \, m_i  }{(j-1)!} 
 \left(\frac{m_i r}{2}\right)^{j-\frac{1}{2}} \, K_{j-\frac{1}{2}} (m_i r) .
\end{split}
\eeq
Putting together the expressions for $A(r)$ and $B(r)$ we can write the metric \eqref{metricB=0} in explicit form,
\beq
\begin{split}
\label{metricB=0-polydeg}
\rd s^2  = & - \bigg[ 1-\frac{2GM}{r} +\frac{8 G M }{\sqrt{\pi} r } \sum_{i=1}^{n} \sum_{j=1}^{\alpha_i} 
\frac{\mathcal{A}_{i,j}}{(j-1)!} 
 \left(\frac{m_i r }{2} \right)^{j+\frac{1}{2}} K_{j-\frac{5}{2}} (m_i r)
 \bigg] 
\exp \bigg[ - \frac{4 G M}{ \sqrt{\pi}}  \sum_{i=1}^{n} \sum_{j=1}^{\alpha_i}  \frac{   \mathcal{C}_{i,j}  m_i  }{(j-1)!} 
 \left(\tfrac{m_i r}{2}\right)^{j-\frac{1}{2}} \, K_{j-\frac{1}{2}} (m_i r) 
 \bigg] \rd t^2 
\\
&
+ \bigg[ 1-\frac{2GM}{r} +\frac{8 G M }{\sqrt{\pi} r } \sum_{i=1}^{n} \sum_{j=1}^{\alpha_i} 
\frac{\mathcal{A}_{i,j}}{(j-1)!} 
 \left(\frac{m_i r }{2} \right)^{j+\frac{1}{2}} K_{j-\frac{5}{2}} (m_i r) 
\bigg] ^{-1} \rd r^2 + r^2 \rd \Omega^2
.
\end{split}
\eeq


\section{Summary and conclusions}
\label{Sec8}

In the present work, we studied static spherically symmetric solutions of the effective field equations~\eq{nlEE2} sourced by an effective energy-momentum tensor containing the smeared delta source. In particular, we discussed the role played by the effective pressure components, several possible ways of defining them, and their consequences to the solution. Throughout the analysis, we kept the effective delta source as general as possible, relating our results to certain generic properties of such sources; this guarantees the generality of the considerations and their applicability to a broad family of models defined by form factors $f(-k^2)$ that grow at least as fast as $k^4$. A summary of the results can be formulated as follows.

If $p_r = p_\th = 0$, the system of field equations is over-determined; nevertheless, it can admit an approximate solution in the regimes of small and large values of $r$. The relevance of this solution is its correspondence to the case in which the original source is a Dirac delta function. Indeed, recalling that the equivalent solution in general relativity would be the Schwarzschild one~\cite{Balasin:1993fn}, it follows that (under the approximations considered) the solution with $p_r = p_\th = 0$ can be regarded as an analogous of Schwarzschild for nontrivial form factors.

On the other hand, the effective field equations can be solved in the whole spacetime if one considers nontrivial effective pressures $p_r$ and $p_\th$. However, there is considerable arbitrariness in the choice of pressure components if one considers equations of state without a guiding physical motivation. In Sec.~\ref{Sec4}, we discussed some possibilities  of equations of state and the solutions associated. In particular, we argued that the choice $p_r = -\rho_{\rm eff}$ makes the shift function $B(r)$ be identically zero and, as a consequence, for large $r$, the resultant metric does not recover the modified Newtonian potential. We also showed that the Newtonian-limit behavior can be matched with the equation of state $p_r = \left[  A(r) - 1 \right] \rho_{\rm eff}$. The explicit solution of the effective field equations with this equation of state was obtained for effective sources originating from local and nonlocal higher-derivative gravity models, as summarized in Table~\ref{TabelaSol}. 

\begin{table}[h]				
\begin{tabular}{c|c|c|cl}
\hline
 $f(z)$ & $A(r)$ & $B(r)$ & Sec. \\
\hline &&&&
\\
 $e^{- z/{\mu^2}}$ & $1 -\frac{2 G M }{r} \, \text{erf}\left(\frac{\mu  r}{2}\right)
+\frac{2 G M \mu }{\sqrt{\pi }} \, e^{-\frac{1}{4} \mu ^2 r^2}
$ & $ -\frac{2 G \mu  M }{\sqrt{\pi }} \, e^{-\frac{1}{4} \mu ^2 r^2}$ &  \ref{Sec7.1}\\
&&&&
\\
\hline
 & & &\\
$\frac{(m_1^2 - z) (m_2^2 - z) }{m_1^2 m_2^2}$ & $1 - 2GM  \frac{ m_1 m_2}{m_2^2 - m_1^2} \left[ m_2 Z(m_1 r)  - m_1 Z(m_2 r) \right]$ & $ -\frac{2 G M m_1 m_2 }{m_2^2 - m_1^2} \left( m_2 e^{-m_1 r} -  m_1 e^{-m_2 r} \right) $ & \ref{Sec7.15}\\
&&&&
\\
\hline &&&&
\\
$\frac{(m^2 - z)^2 }{m^4}$ & $1 - \tfrac{2 G M }{r} \left[  1 - \left( 1 + m r + \tfrac{1}{2} m^2 r^2 \right)  e^{-m r} \right]$ & $ - G M m (1 + m r) e^{- m r}$ & \ref{Sec7.15}\\
&&&&
\\
\hline
 & & & &\\
   $\prod_{i=1}^N \frac{m_i^2 - z}{m_i^2}$ & 
   $1 - \frac{2GM}{r}\; \left[ 1 - \sum_{i=1}^N C_{i} \, \, (1+m_i r)e^{-m_i r} \right] $ & $ -2GM \sum_{i=1}^N C_{i} \, m_i \, e^{- m_i r}$ & \ref{Sec7.2}\\
   &&&&
\\
\hline  
& & & &\\  
$\prod_{i=1}^{n} \left( \frac{m_i^2 - z}{m_i^2}\right)^{\alpha_i} $ & $1-\frac{2GM}{r} +\frac{8 G M }{\sqrt{\pi} r } \sum_{i=1}^{n} \sum_{j=1}^{\alpha_i} 
\frac{\mathcal{A}_{i,j}}{(j-1)!} 
 \left(\frac{m_i r }{2} \right)^{j+\frac{1}{2}} K_{j-\frac{5}{2}} (m_i r) $ & $ - \frac{4 G M}{ \sqrt{\pi}}  \sum_{i=1}^{n} \sum_{j=1}^{\alpha_i}  \frac{   \mathcal{C}_{i,j} \, m_i  }{(j-1)!} 
 \left(\frac{m_i r}{2}\right)^{j-\frac{1}{2}} \, K_{j-\frac{1}{2}} (m_i r) $ & \ref{Sec7.3}\\
 &&&&
\\
\hline  
\end{tabular}
\caption{Solutions obtained in this work for several form factors $f(z)$; the shift function $B(r)$ presented follows from the equations of state of Sec.~\ref{Sec4.4}, namely, $p_r = [A(r)-1]\rho_{\rm eff}$. The function $Z(x)$ is defined in Eq.~\eq{ZndPoly}, while the coefficients $C_i$, $\mathcal{C}_{i,j}$, and $\mathcal{A}_{i,j}$ are defined, respectively, in Eqs.~\eq{coi}, \eq{G_coeff_C}, and~\eq{relationAC}.  
The last column refers to the section where the solution is discussed.}
\label{TabelaSol}
\end{table} 

We remark that all the solutions found, regardless of the choice of the effective pressures, present a black hole mass gap. In other words, the metric does not have an apparent horizon if the mass is smaller than a critical value.

A significant part of the work was dedicated to studying the regularity of the solutions and how they compare with the Newtonian limit metric. We showed that if the form factor $f(-k^2)$ of the model grows at least as fast as $k^4$, there are choices of equations of state that result in the curvature invariants being singularity-free. The same result is generalized to guarantee the regularity of sets of curvature invariants containing covariant derivatives; however, in this case, the behavior of the form factor should be improved accordingly. The whole set of invariants that are polynomial in curvature tensors and their covariant derivatives can be regular if $f(-k^2)$ grows faster than any polynomial (i.e., in some nonlocal gravity models).

Regarding the comparison with the nonrelativistic limit, we showed that the solutions related to the equation of state $p_r = \left[  A(r) - 1 \right] \rho_{\rm eff}$ match the Newtonian-limit solutions not only in the regime of large $r$ but also for small $r$ (if the effective source is regular). This means that the curvature invariants show the same leading behavior in these regions; however, higher-order derivative invariants may differ at $r=0$.

We highlight that even though our discussion was centered on smeared sources from higher-derivative gravity models, the same effective delta sources can originate from other quantum gravity approaches, e.g., noncommutative geometry, generalized uncertainty principle models, and string theory~\cite{Cadoni:2022chn,Cadoni:2023nrm,Akil:2022coa,Mureika:2010je,Gaete:2010sp,dirty,Nicolini:2005vd,Isi:2013cxa,Tseytlin:1995uq,Modesto:2010uh,Nicolini:2019irw,Nicolini:2012eu,Knipfer:2019pgi}. Hence, the general results derived here can also be applied in other frameworks.

It has yet to be determined if the singularities of the classical solutions of general relativity can be resolved at a more fundamental quantum level. Nevertheless, our results may suggest that they can be regularized by higher-derivative structures in the gravitational action. Indeed, the higher-derivative modifications of the Schwarzschild metric presented here are nonsingular and display the same qualitative behavior of the solutions reported in~\cite{Holdom:2002xy}. Although a rigorous analysis of the complete field equations in the case of models with six or more derivatives is still pending, this work can motivate and be useful for further research on this important topic.


\section*{Acknowledgments}

This work was supported by the Basic Research Program of the Science, Technology, and Innovation Commission of Shenzhen Municipality (grant no.\ JCYJ20180302174206969).  B.L.G. is supported by Primus grant PRIMUS/23/SCI/005 from Charles University.


\appendix

\begin{appendices}

\renewcommand{\thesubsection}{\thesection.\arabic{subsection}}  

\section{Case $p_r = \left[   \tfrac{\mu  r}{2} A(r) - 1 \right]  \rho_{\rm eff}$}
\label{App1}

For the sake of completeness, here we discuss the case of the equation of state that comes from the choice $\si(r) \propto r \rho_{\rm eff}(r)$ in the framework of Sec.~\ref{SecPresGen}. This example is interesting for at least three reasons: \textit{i}) The pressure components are regular, but they do not vanish as $r\to 0$, \textit{ii}) the equation of state introduces an odd power of $r$ in the effective pressures, and \textit{iii}) the same equation of state was used in Ref.~\cite{dirty} in the context of noncommutative geometry with the Gaussian effective source~\eq{gauss}, motivating a comparison of that solution with our metric~\eq{metricB=0-exp}.

Hence, let us choose the equation of state
\beq
\n{EOS21}
p_r = \left[  \frac{\mu  r}{2} A(r) - 1 \right] \rho_{\rm eff},
\eeq
where $\mu$ is a parameter with dimension of mass, which could be linked to the massive parameters of the higher-derivative model. In this case, 
the conservation equation~\eq{Conserva} defines
\beq
p_\th = \frac{\mu r}{8}   \left\lbrace   3 r \rho_{\rm eff} A^\prime + \left[ (6 + r B^\prime ) \rho_{\rm eff} + 2 r \rho_{\rm eff}^\prime  \right] A \right\rbrace    - \frac{1}{2} \left( 2 \rho_{\rm eff} + r \rho_{\rm eff}^\prime \right)   .
\eeq
Both pressures satisfy the criteria (1b) and (2b) of Sec.~\ref{Sec4.4}---but not (3b).

An advantage of using~\eq{EOS21} is the simplicity of the expression for $B(r)$, since it makes Eq.~\eq{EqGrr} become the same equation defining the effective mass $m(r)$. Therefore, together with the condition of matching Minkowski at  infinity, it gives
\beq
\n{BBM}
B(r) = G \mu \left[ m(r) - M \right] .
\eeq

Notice, however, that since the effective pressures do not vanish at the origin, the behavior of this solution near $r=0$ might be very different from the ones presented in the main part of the paper (see Sec.~\ref{Sec4}). As a simple example, recall that the mass function $m(r)$ is an odd function 
for nonlocal models defined by form factors that grow faster than any polynomial [see Eq.~\eq{m-odd} and the related discussion]; hence, aside from a constant, the series representation of~\eq{BBM} has only \emph{odd} powers of $r$, whereas, \emph{for the same models}, the solutions~\eq{bsol2}, \eq{bsolalpha}, and~\eq{SolB-A} have only \emph{even} powers. Let us remind that the absence of odd powers in the series expansion of the metric is crucial to the regularity of scalars containing derivatives of curvature tensors~\cite{Nos6der,Giacchini:2021pmr}, as discussed also in Sec.~\ref{Sec6}. 

The presence of odd powers of $r$ in the series of~\eq{BBM} is generally expected for higher-derivative models. In fact, from the small-$r$ behavior of the mass function in Eq.~\eq{limM2} we obtain the expansion of~\eq{BBM} around $r=0$,
\beq
\n{limB4}
B(r) = G \mu M \left[ -1 + \frac{4\pi \rho_{\rm eff}(0)}{3M}  r^3\right]  + O(r^4),
\eeq
if  $f(-k^2) \sim k^4$ (or faster) for large $k$.
Thus, for the higher-derivative gravity models considered in this paper, one can always guarantee the presence of the term $r^3$ in the series expansion of $B(r)$ if the equation of state~\eq{EOS21} is used. Moreover, the term linear in $r$ is always absent, which ensures the regularity of curvature scalars \emph{without covariant derivatives} (see Sec.~\ref{Sec6}).

Taking into account the relation between the asymptotic behavior of $f(-k^2)$ and the lowest odd power of $r$ in the series of $A(r)$ [see Eq.~\eq{AprideG}], the Eq.~\eq{boquisErre} guarantees that the scalar $\Box R$ is singular at $r=0$ for any model with $f(-k^2) \sim k^6$ (or faster) for large $k$ (e.g., eighth- and higher-derivative gravity), if the equation of state~\eq{EOS21} is used. In fact, the explicit calculation gives
\beq
\n{AppBoxR}
\Box R =  -\frac{32 \pi G \mu  \rho_{\rm eff}(0)}{r} + O(r^0).
\eeq
In the terms of~\cite{Giacchini:2021pmr}, this happens because, while the first odd power in the series expansion of $g_{rr}$ is at least $O(r^5)$ [see Eq.~\eq{AprideG}], for $g_{tt}$ it is for sure $O(r^3)$, see Eq.~\eq{limB4}. 

Only in the case of $f(-k^2) \sim k^4$ (like in sixth-derivative gravity) $\Box R$ might be regular~\cite{Giacchini:2021pmr}. This occurs because $A(r)$ contains a term $a_3 r^3$ in the series expansion [$a_3 = -2 \pi G \rho_{\rm eff}^\prime(0) \neq 0$], which opens a possibility for $\Box R$ be regular if
\beq
- 40 a_3 - 32 \pi G \mu  \rho_{\rm eff}(0) = 0,
\eeq
as it can be deduced from the comparison of~\eq{AppBoxR} and~\eq{boquisErre}. One can also verify that under this exceptional condition, the other scalars with two covariant derivatives (such as $(\nabla_\la R_{\mu\nu\al\beta})^2$ and $R_{\mu\nu} \Box R^{\mu\nu}$) are also regular.

Last but not least, in what concerns the comparison with the Newtonian limit, we notice that the higher powers of $r$ entering the equations of state considered in this Appendix make the effective pressures vanish more slowly than the ones of Sec.~\ref{Sec4.4}, as $r$ increases. Thus, the solution in the regime of large $r$ might likely converge to the Newtonian-limit metric in a more slow way.

\begin{figure}[t]
\centering
\begin{subfigure}{.5\textwidth}
\centering
\includegraphics[scale=0.7]{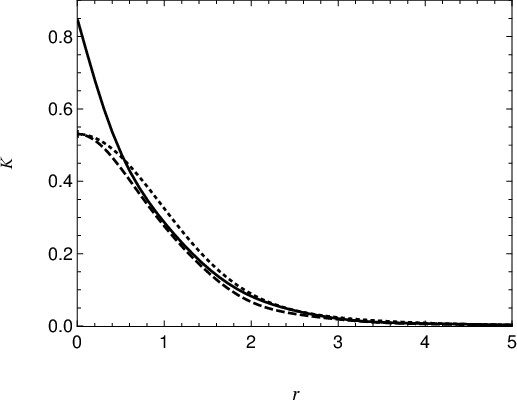}
\caption{\small Kretschmann scalar.}
\label{FigAppA}
\end{subfigure}%
\begin{subfigure}{.5\textwidth}
\centering
\includegraphics[scale=0.7]{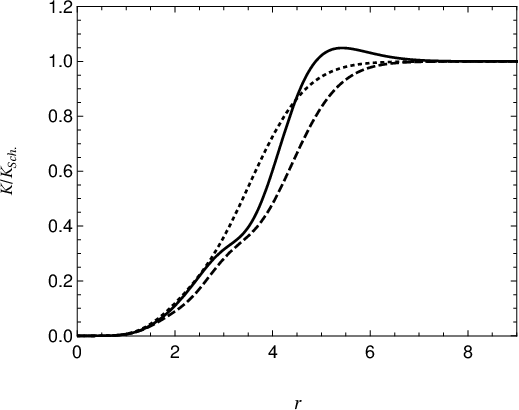}
\caption{\small Ratio $K/K_{\text{Sch.}}$.}
\label{FigAppB}
\end{subfigure}
\caption{\small Kretschmann scalar $K$ and the ratio $K/K_{\text{Sch.}}$ with $K$ evaluated at the metrics~\eq{metricB=appendix} (solid line),
\eq{metricB=0-exp} (dashed line), and at the Newtonian-limit metric~\eq{linme} (dotted line) with potential~\eq{maz_sqn}.}
\end{figure}

As a concrete example, let us consider the case of the Gaussian effective source~\eq{gauss}. The main choice of equation of state used in this paper yields the metric~\eq{metricB=0-exp}, while the choice~\eq{EOS21} results in
\beq
\begin{split}
\label{metricB=appendix}
\rd s^2 = &- \left[1 -\frac{2 G M }{r} \, \text{erf}\left(\frac{\mu  r}{2}\right)
+\frac{2 G M \mu }{\sqrt{\pi }} \, e^{-\frac{1}{4} \mu ^2 r^2} \right] 
\exp \left\{
-G  \mu M \left[1 - \text{erf}\left(\frac{\mu  r}{2}\right) + \frac{\mu r }{\sqrt{\pi }} \,  e^{-\frac{1}{4} \mu ^2 r^2} \right]
\right\} \rd t^2 
\\
&
+ \left[ 1 -\frac{2 G M }{r} \, \text{erf}\left(\frac{\mu  r}{2}\right)
+\frac{2 G M \mu }{\sqrt{\pi }} \, e^{-\frac{1}{4} \mu ^2 r^2} \right]^{-1} \rd r^2 + r^2 \rd \Omega^2
.
\end{split}
\eeq
As anticipated, the metric~\eq{metricB=appendix} coincides with the one obtained in~\cite{dirty} through the replacement $\mu = 1/\sqrt{\th}$ and a straightforward manipulation involving the lower incomplete gamma function $\ga(a,x)$ using the formula 
\beq
\ga\left( \tfrac{3}{2} , \tfrac{\mu^2 r^2}{4} \right) 
= \frac{1}{2}  \left[ \sqrt{\pi} \, {\rm erf} \left(\frac{\mu r}{2}\right) - \mu r \, e^{-\frac14 \mu^2 r^2} \right] .
\eeq 

In Fig.~\ref{FigAppA}, we plot the Kretschmann scalar for the metrics~\eq{metricB=0-exp}, \eq{metricB=appendix}, and for the Newtonian solution with the potential~\eq{maz_sqn}. It shows that the solution~\eq{metricB=appendix} differs significantly from the others as $r \to 0$. In Fig.~\ref{FigAppB}, we compare the Kretschmann scalar evaluated at all the three solutions mentioned above with 
the one for the Schwarzschild solution, Eq.~\eq{KSch}. 
The three solutions satisfy $K/K_{\text{Sch.}} \to 1$ as $r\to\infty$, i.e., in the IR they tend to the Schwarzschild solution, but the presence of the higher powers of $r$ in the equation of state~\eq{EOS21} makes the solution~\eq{metricB=appendix} converge slightly more slowly.

\end{appendices}



\end{document}